\documentclass[a4paper]{aa}  

\usepackage{graphicx}
\usepackage{subcaption}
\usepackage{lscape}
\usepackage[varg]{txfonts}

\usepackage{color}
\usepackage{natbib,twoopt}
\usepackage[hyphenbreaks]{breakurl}
\usepackage[breaklinks]{hyperref}      
\bibpunct{(}{)}{;}{a}{}{,}             
\definecolor{cobalt}{rgb}{0.06, 0.2, 0.65}
\hypersetup{
  colorlinks,
  citecolor=cobalt,
  linkcolor=[rgb]{0.8, 0.2, 1.0},
  urlcolor=cobalt,
}
\makeatletter
  \newcommandtwoopt{\citeads}[3][][]{\href{http://adsabs.harvard.edu/abs/#3}%
    {\def\hyper@linkstart##1##2{}%
     \let\hyper@linkend\@empty\citealp[#1][#2]{#3}}}
  \newcommandtwoopt{\citepads}[3][][]{\href{http://adsabs.harvard.edu/abs/#3}%
    {\def\hyper@linkstart##1##2{}%
     \let\hyper@linkend\@empty\citep[#1][#2]{#3}}}
  \newcommandtwoopt{\citetads}[3][][]{\href{http://adsabs.harvard.edu/abs/#3}%
    {\def\hyper@linkstart##1##2{}%
     \let\hyper@linkend\@empty\citet[#1][#2]{#3}}}
  \newcommandtwoopt{\citeyearads}[3][][]%
    {\href{http://adsabs.harvard.edu/abs/#3}
    {\def\hyper@linkstart##1##2{}%
     \let\hyper@linkend\@empty\citeyear[#1][#2]{#3}}}
\makeatother

\begin{document}

   \title{Massive black hole formation in Population III star clusters}


   \author{B. Reinoso
          \inst{1,2,}
          \and
          M. A. Latif\inst{3}
          \and
          D.R.G. Schleicher\inst{4,5}
          }

   \institute{Department of Physics, Gustaf H\"allstr\"omin katu 2, FI-00014, University of Helsinki, Finland \\
              \email{bastian.reinoso@helsinki.fi}
         \and
             Universit\"at Heidelberg, Zentrum f\"ur Astronomie, Institut f\"ur Theoretische Astrophysik, Albert-Ueberle-Str. 2, 69120 Heidelberg, Germany
         \and
           Physics Department, College of Science, United Arab Emirates University, PO Box 15551, Al-Ain, UAE\\
             \email{latifne@gmail.com}
        \and
            Dipartimento di Fisica, Sapienza Università di Roma, Piazzale Aldo Moro 5, 00185 Rome, Italy\\
            \email{dominik.schleicher@uniroma1.it}
         \and
            Departamento de Astronom\'ia, Facultad Ciencias F\'isicas y Matem\'aticas, Universidad de Concepci\'on, Av. Esteban Iturra s/n, Concepci\'on, Chile            
             }

   \date{Received March 26, 2025; accepted --}

 
  \abstract
   {The James Webb Space Telescope (JWST) has revealed a population of active galactic nuclei (AGNs) that challenge existing black hole (BH) formation models. These newly observed BHs seem overmassive compared to the host galaxies and have an unexpectedly high abundance. Their exact origin remains elusive.}
   {The primary goal of this work is to investigate the formation of massive BH seeds in dense Population III (Pop~III) star clusters.}
   {Using a cosmological simulation of Pop~III cluster formation, we present models for the assembly and subsequent evolution of these clusters. The models account for background gas potential, stellar collisions and associated mass loss, gas accretion, stellar growth, their initial mass function (IMF), and subsequent star formation. We conduct $N$-body simulations of these models over a span of 2 million years.}
   {Our results show that BHs of $>400$~M$_\odot$ are formed in all cases, reaching up to $\sim 5000$~M$_\odot$ under optimistic yet reasonable conditions, and potentially exceeding $10^4$~M$_\odot$ provided that high accretion rates of $10^{-3}$~M$_\odot$~yr$^{-1}$ onto the stars can be sustained.} 
   {We conclude that massive BHs can be formed in Pop III stellar clusters and are likely to remain within their host clusters. These BHs may experience further growth as they sink into the galaxy' s potential well. This formation channel should be given further consideration in models of galaxy formation and BH demographics.}


   \maketitle
   
%
\nolinenumbers
\section{Introduction}
During the past two decades, a large number of high redshift quasars have been discovered, with more than 300 at $z>5.7$ \citep{Fan2006,Mortlock2011,Wu15Nature,Banados2018,Reed2019,Onoue2019,Banados2021,Wang21}. These observations imply the existence of supermassive black holes (SMBHs) with 10$^9$-10$^{10}$~M$_\odot$ when the Universe was less than 1~Gyr old. In recent years, the JWST has pushed these limits to even higher redshifts and lower masses. 


Recent JWST surveys have unveiled a large population of faint AGN candidates at $z \geq$ 4 \citep{Kok23,Furt23,Lamb23,Lars23,Gould23,Maio23a,Bogdan24,Kovacs24,Mat24,Koko24,Greene24,Napolitano24}. Among the most intriguing discoveries are very compact, reddish objects that exhibit broad line emission features in their spectra, known as the little red dots \citep[LRDs;][]{Mat24,Greene24,Koc24,Mai24}. These newly discovered BHs have masses of a few times $10^5-10^7$~M$_{\odot}$, less massive than quasars at the same redshifts but, interestingly, similar in mass at the low end to those expected for direct collapse black holes at birth \citep[DCBHs;][]{Hosokawa13,tyr17,L22N,pat23}. They seem overmassive relative to their host galaxies when compared to the local $\rm M_{BH} - M_{star}$ relation \cite[also see][]{Lupi2024,King24,Lambrides24,Rusakov25,Naidu25}) and their bolometric luminosities range from $\rm 10^{44}- 10^{45} ~erg/s$. They reside in galaxies with sub-solar metallicities.  These discoveries provide key insights into the AGN population, the co-evolution of AGNs with their host galaxies, BH seeding, and growth mechanisms. The cosmic abundance of these recently observed AGN is $\rm \sim 10^{-5}\ Mpc^{-3}\ mag^{-1}$, about two orders of magnitude higher than the faintest UV selected quasars observed at such redshifts \citep{Mat24,Koko24,Greene24}. 


On the theoretical side, several possibilities exist to explain the observed SMBHs.  The most straightforward path for BH seed formation in the early Universe is the gravitational collapse of massive Pop~III stars. Despite the uncertainties in the Pop~III IMF, the general consensus is to consider BH seeds of 100~M$_\odot$. If the $z\gtrsim7$ quasars are descendants of these light seeds, they must have grown uninterruptedly at the Eddington rate. Even for some newly discovered JWST AGNs, super-Eddington accretion would be necessary for such low initial BH masses \citep{Bogdan24,Kovacs24, Lupi2024, Napolitano24}. However, numerical simulations have shown that the birthplaces of these BHs are rapidly depleted of gas due to supernova (SN) feedback, that these BH seeds tend to wander within their host galaxies due to inefficient dynamical friction, and that they rarely encounter cold, dense gas from which to accrete mass. Moreover, they must compete with star formation for the available cold gas \citep{JohnsonBromm2007,Alvarez2009,Pacucci2017,Smith2018,latif20b}. Natural explanations for the observation of supermassive black holes at $z>9$ have been put forward e.g. by \citet{Schneeider2023}.

Some of the hurdles faced by this formation channel are easily overcome by the DCBH scenario, which produces BH seeds of $10^4 - 10^5$~M$_\odot$, born in gas-rich environments \citep[see e.g.][]{Regan2009,Regan2009b,Latif13c,Latif2016,Chon2018,Wise2019,Regan2020b, Chon2020, L22N,Schleicher2023, Saavedra2024,Chon24}. Depending on the strength of the radiative background, the masses can also lie within 10$^3$ to 10$^4$~M$_\odot$ \citep{LAtif2014,Latif2015,Regan2020b,Latif2021}. Many uncertainties remain regarding the abundance expected from the DCBH scenario, with estimates differing by several orders of magnitude \citep{Habouzit2016}. It now seems that even optimistic assumptions about DCBH formation are incapable of accounting for the high abundance of high-$z$ AGNs, assuming that all LRDs are indeed powered by SMBHs. Recent models seem to favor lighter but more abundant seeds \citep{Latif2021,Sassano2021,Trinca24,Regan2024,OBrennan25}.

In light of this new evidence, the exploration of alternative BH formation pathways is worth pursuing. One such mechanism consists of the growth of a central object due to stellar collisions in very dense star clusters \citep{Omukai2008, Katz2015, Sakurai2017, Sakurai2019, Reinoso2018, Reinoso20, Vergara2021,Rantala24}. This formation channel has the advantage of producing more massive seeds than the Pop~III remnants, and they are embedded in massive systems which aid during the sinking to the galactic center \citep{Mukherjee24}. Moreover, these seeds should be more abundant than DCBH seeds, which would alleviate tensions with recent JWST data. Semi-analytic models show that they could contribute quite substantially to explain the SMBH population in the present-day Universe \citep{Liempi2025}. Interestingly, the dense stellar systems in which these seeds should form, with effective radii $\lesssim 1$~pc and masses $M\gtrsim10^6$~M$_\odot$, are now being revealed by JWST observations \citep{Adamo24}. An alternative pathway, in case the stellar evolution is faster than the dynamical evolution of the cluster, consists of the formation of a cluster of stellar mass BHs which may migrate to the center and lead to the formation of a massive BH via gravitational wave emission if the potential of the BH clusters is sufficiently steep, for example due to an external gas inflow \citep{Davies2011,Lupi2014, Kroupa2020, Gaete2024}.





The collision-based channel has become increasingly relevant in recent years. From an analysis of the observed parameter space of Nuclear Star Clusters (NSCs) with and without SMBHs, \citet{Escala2021} has shown that NSCs with SMBHs exist in the region where the collision time in the cluster is comparable to the cluster age, while no SMBHs have been detected in clusters where the age is shorter than the collision time. Via numerical simulations, \citet{Vergara23} have tested this hypothesis and have shown that the efficiency to form a central object via collisions increases significantly when the cluster mass is close to a critical mass scale, which depends on the number and mean mass of stars as well as the size and age of the cluster. This result has been confirmed in a more extensive analysis that includes a larger suite of numerical simulations, such as e.g. \citet{ArcaSedda23,ArcaSedda24}, as well as a comparison with observational data from the Local Universe \citep{Vergara2024}.

In the presence of gas, it is quite likely that the mass of the resulting objects will be further enhanced, given that both the potential steepens and  the crossing time shortens, increasing the possible frequency of collisions \citep{Reinoso20}. In addition, the dynamical friction due to the gas creates a more dissipative system, and also the accretion of gas onto the stars enhances the probability of collisions  \citep{Boekholt2018, Tagawa2020, Schleicher2022}. These effects have also been demonstrated in simulations with gas \citep{Reinoso23, Saavedra2024} and are also implemented in the latest version of Monte-Carlo codes for the modeling of star clusters \citep{Giersz2024}.

In this paper, we present a simulation performed with the \textsc{ENZO} code \citep{ENZO} to model the formation and initial evolution of a primordial star cluster in a cosmological context. To explore its longer-term evolution, this cluster is then further evolved using the Astrophysical MUlti-purpose Software Environment (AMUSE) framework \citep[e.g.][]{Portegies2018}, considering physically motivated recipes  for accretion and collisions and the mass-radius relation for protostars.

We describe our numerical methodology in Section~\ref{sec:Methods} and present the main results in Section~\ref{sec:results}. Our conclusions and discussion are provided in Section ~\ref{sec:Discussion_and_conclusions}. Additionally, Appendix~\ref{sec:orb_params_colls} includes a short compilation of orbital parameters that led to stellar collisions in our simulations, which can aid in further studies and improve our understanding of mass loss during stellar collisions.



\section{Methods}
\label{sec:Methods}

%

We simulate the formation of a Pop III cluster in a pristine dark matter halo by performing a cosmological radiation hydrodynamical simulation and coupling its output with the \textsc{AMUSE} framework to conduct N-body simulations, enabling us to follow the evolution of the Pop III cluster. The setup of our cosmological simulation and the initial conditions for \textsc{AMUSE} are described below.

\subsection{Cosmological simulation setup}

Numerical simulations are conducted with the cosmological hydrodynamics open source code  \textsc{ENZO} \citep{ENZO} which uses an adaptive mesh refinement approach to the cover a large range of spatial scales: from tens of Mpc  down to sub parsec scales.  \textsc{ENZO} employs a  3rd order piece wise parabolic solver  for hydrodynamics, a $N$-body particle-mesh to compute DM dynamics and multigrid Poisson solver for self-gravity calculations. Our simulations start  at $z$=200 with cosmological initial conditions generated from Gaussian random density perturbations with MUSIC  \citep{Hahn11}  and use the following cosmological parameters from the \textit{Planck} data: $\Omega_{\mathrm{M}}=$ 0.3089, $\Omega_{\Lambda}=$ 0.691, $\Omega_{\mathrm{b}} = $ 0.0486, $h =$ 0.677, and $n =$ 0.967 \citep{Planck16}.  They are performed in a volume of $\rm (50 ~Mpc/h)^3$ and were initially ran down to $z$=7 with coarse DM  resolution of $\rm 512^3$  which yields a DM mass resolution of  $\rm 10^8$~M$_\odot$, sufficient enough to resolve the target halo of $\rm 1.6 \times 10^{12}$~M$_\odot$ at $z$=7  by 10,000 DM particles. To capture such a rare halo, we boosted the value of $\sigma_8$ to 1.2 as this approach has been used in previous works, see \cite{Hirano2017}. Using the rockstar halo finder \citep{RS}, we identify the DM particles belonging to the halo in its Lagrange volume and trace them back to the initial redshift. We then add one refinement level in a region twice the Lagrange radius of the halo and run a zoom-in simulation. We keep repeating these steps, every time adding one additional refinement level in the region twice the Lagrange volume of the target halo until we achieve a DM resolution of a few thousand solar masses ($\sim$ 3600~M$_\odot$) in the zoomed in region\footnote{see https://github.com/jwise77/enzo-mrp-music for details}. This approach allows us to resolve minihalos of $\rm 10^5$~M$_\odot$.  We further employ up to 18 dynamical levels of refinement in the region of interest during the course of simulation, which enable us to resolve the gravitational  collapse down to the scales of about 0.01 pc. Our refinement criteria are based on the baryonic overdensity, DM mass resolution and  the Jeans refinement, for details see \citep{L22,latif20b,L22N}. We resolve the Jeans length by at least 64 cells which is found to be sufficient to resolve turbulent eddies in cosmological simulations \citep{Fed11,Latif13c}. Similar to \cite{L22N}, cold flows drive supersonic turbulence that prevents star formation in the halo until it reaches a mass of a few times $\rm 10^7~M_{\odot}$ at z=30 (see Fig. \ref{fig:velocity_profile} which shows the comparison of turbulent and radial infall velocites at different times) and catastrophic gravitational collapse gets triggered. Dense cold streams of gas penetrate the halo and drive turbulent motions of a few 10 km/s, as shown in Fig. \ref{fig:large_scale_halo}.

\begin{figure}[h]
   \centering
   \includegraphics[width=9cm]{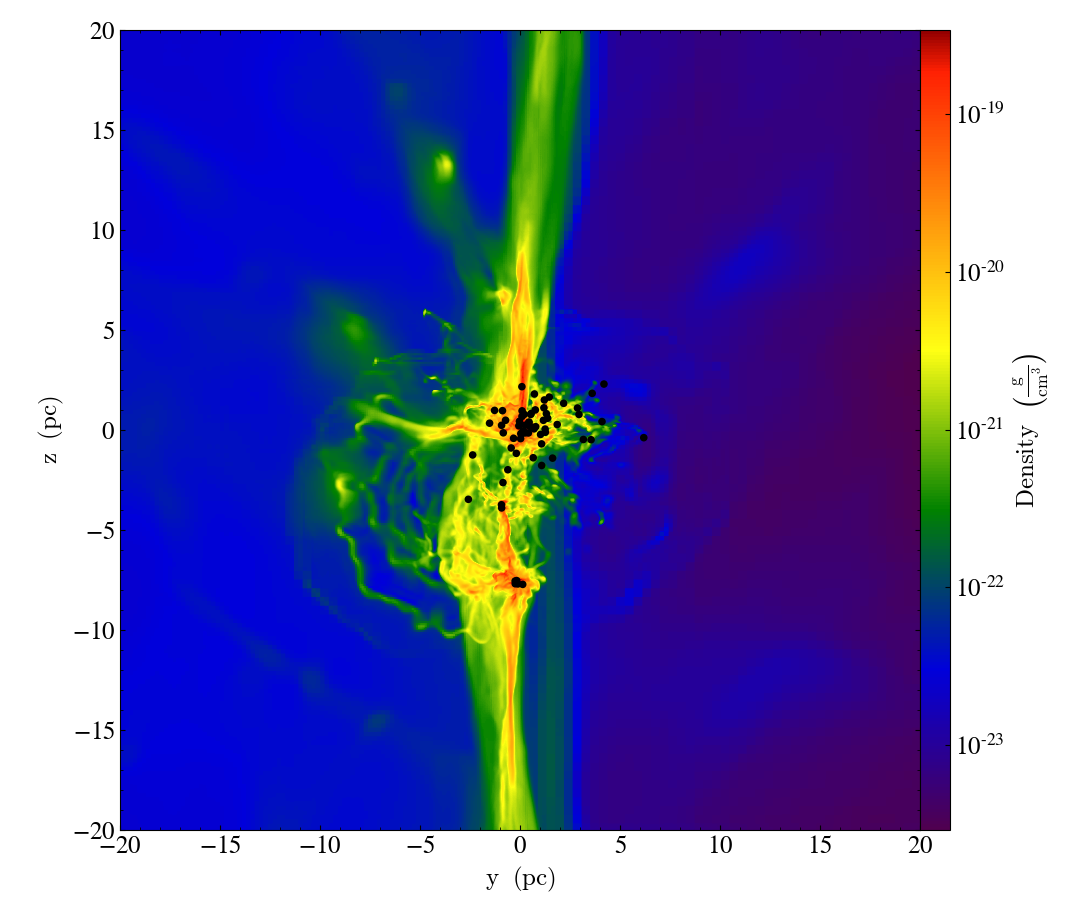}
      \caption{Snapshot of an infant Pop~III star cluster obtained with \textsc{ENZO}. We utilize this data for constructing models of the assembly and further evolution of embedded Pop~III star clusters.
              }
         \label{fig:snapshot_ENZO}
   \end{figure}

At this stage, we turn on sink particles which represent Pop III stars in our simulations and are typically formed at densities of $10^{-18} \rm g/cm^3$. Our recipe for sink formation is based on SmartStars \citep{Regan18b,L22} and sinks are created in  a grid cell when it meets the following criteria: I) at the maximum refinement level, II) at a local minimum of the gravitational potential, III)  has a convergent flow, IV) density is greater than the Jeans density, V) the gas cooling time is shorter than the collapse time.  A sink particle can accrete and gets merged with more massive sink forming in the accretion radius corresponding to 4 cells. The stellar feedback is modeled assuming a black-body spectrum with a temperature of $\rm T_{eff}  = $ 5000 K for $\rm M_{*}  \leq$ 10~M$_\odot$\ \citep{Hosokawa13} and $\rm T_{eff} = 10^{4.759}$ when $\rm M_{*} >$ 10~M$_\odot$\ \citep{Schaerer02}. The radiative transfer scheme, MORAY, is used to trace radiation from stars. The flux is divided into five energy bins: two for dissociating photons (2.0 eV, 12.8 eV) and three for ionizing photons (14.0 eV, 25.0 eV, 200.0 eV), with bin definitions determined using the SEDOP code developed by \citet{Mirocha12}. For  further details, the reader is referred to  \cite{L22}. MORAY is coupled to hydrodynamics and non-equilibrium chemistry solver based on \cite{Abel97}. The chemical model solves the rate equations for nine primordial species: $ \rm H, ~H^+,~ H^-, ~He,~ He^+, ~He^{++},~ H_2, ~H_2^+, ~e^-$ and and includes all relevant processes, such as $\rm H_2$ formation via three-body reactions, $\rm H_2$ cooling as well as  cooling from collisional excitation, collisional ionization, radiative recombination, Bremsstrahlung radiation, CIE cooling, $\rm H_2$ optically thick cooling approximation at densities $\rm \geq 10^{-14}~g/cm^3$ and heating from three-body reactions. We ignore deuterium and related species because HD cooling mainly occurs in relic H II regions or shock-heated gas during major mergers which  boost D$^+$ abundance \citep{Yoshida2007,Maio2007,McGreer2008,Greif2008,Bovino2014}. Therefore, we expect it not to be too relevant at least in massive halos like the one simulated here. We take the simulation output from \textsc{ENZO} at $z=39.4$ when a total of 68 sink particles have formed, containing a total stellar mass of 1497~M$_\odot$ and feed it to AMUSE. The simulation was evolved for 330~Kyrs and a snapshot of the state of the system at this time is shown in Fig.~\ref{fig:snapshot_ENZO}.

\subsection{Models for the evolution of a Pop~III star cluster}


We take the positions, velocities, masses, and accretion rates of Pop III stars, along with the mass of the parent gas cloud and the gas inflow rate into the cluster from the \textsc{ENZO} simulation. These are then used to perform $N$-body simulations within the AMUSE framework. We model six different Pop III star clusters based on the \textsc{ENZO} simulation output, which are listed in Table \ref{tab:sim_set}. To account for the effect of gas, we model it as a background analytic potential and increase its strength in proportion to the gas inflow rate. For unresolved star formation, we assume three different star formation efficiencies: 10\%, 20\%, and 30\%, to decide the total stellar mass. We also include gas accretion onto the stars, using mass-radius relations appropriate for Pop III stars (see Section~\ref{sec:MR_relation}), and consider stellar collisions and the associated mass loss (see Section~\ref{sec:collisions_mass_loss}). In total, we perform 18 simulations and evolve them for 2 Myr, corresponding to the lifetime of the most massive Pop III star. The simulations were carried out using the AMUSE\footnote{https://github.com/amusecode/amuse} framework \citep[see][]{AMUSE_Portegies09,AMUSE_Portegies13,AMUSE_Pelupessy13,Portegies2018}, by coupling the pure $N$-body code \textsc{PH4} \citep{McMillan96} with an analytic background potential using the {\small BRIDGE} method \citep{Fujii2007}. These processes are described in more detail below.
   


   \subsection{Mass-radius relation}
   \label{sec:MR_relation}

\begin{figure}[h]
   \centering
   \includegraphics[width=9cm]{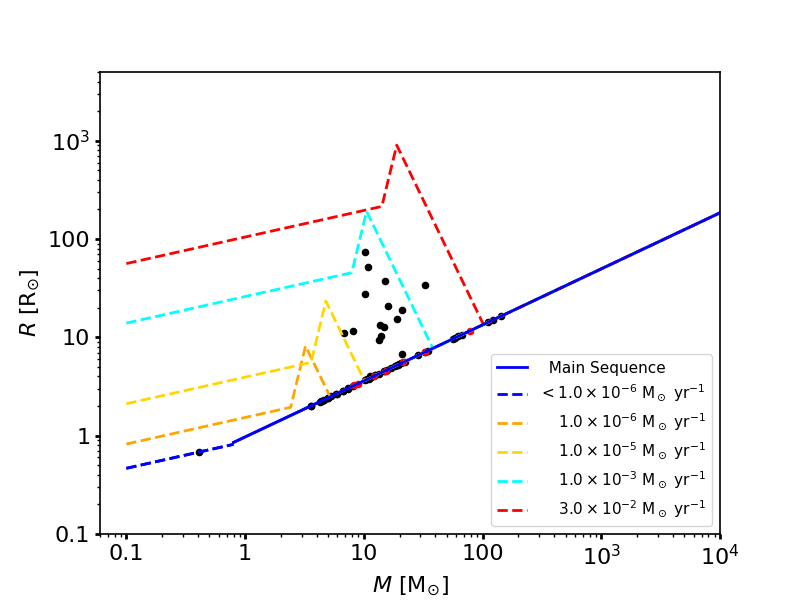}
      \caption{Mass-radius relations for accreting Pop~III stars. Dashed lines show the relations followed by protostars accreting at different rates, whereas the solid blue line is the relation followed by main sequence stars. The filled circles show the mass and radius for each particle at the beginning of our simulations. The stellar radii for black circles were obtained with the accretion rates provided by \textsc{ENZO}, whereas the stellar radii for red circles were obtained assuming $\dot{m}=0.0$~M$_\odot$/yr.
              }
         \label{fig:mass-radius}
   \end{figure}

To account for stellar growth, we use the mass-radius relation based on the work of \cite{Hosokawa09} and \cite{Hosokawa12,Hosokawa13}. This relation is applicable to accreting protostars and main-sequence stars of primordial composition. For a detailed description, we refer the reader to \citet{Reinoso23}. In Fig.~\ref{fig:mass-radius}, we present the mass-radius relation for different mass accretion rates, along with the positions of the stars at the start of our simulations, which are taken from the \textsc{ENZO} run. We note that all stellar collisions in our simulations involve main-sequence stars only.

   
   
\subsection{The background potential}
\begin{figure}[h]
   \centering
   \includegraphics[width=9cm]{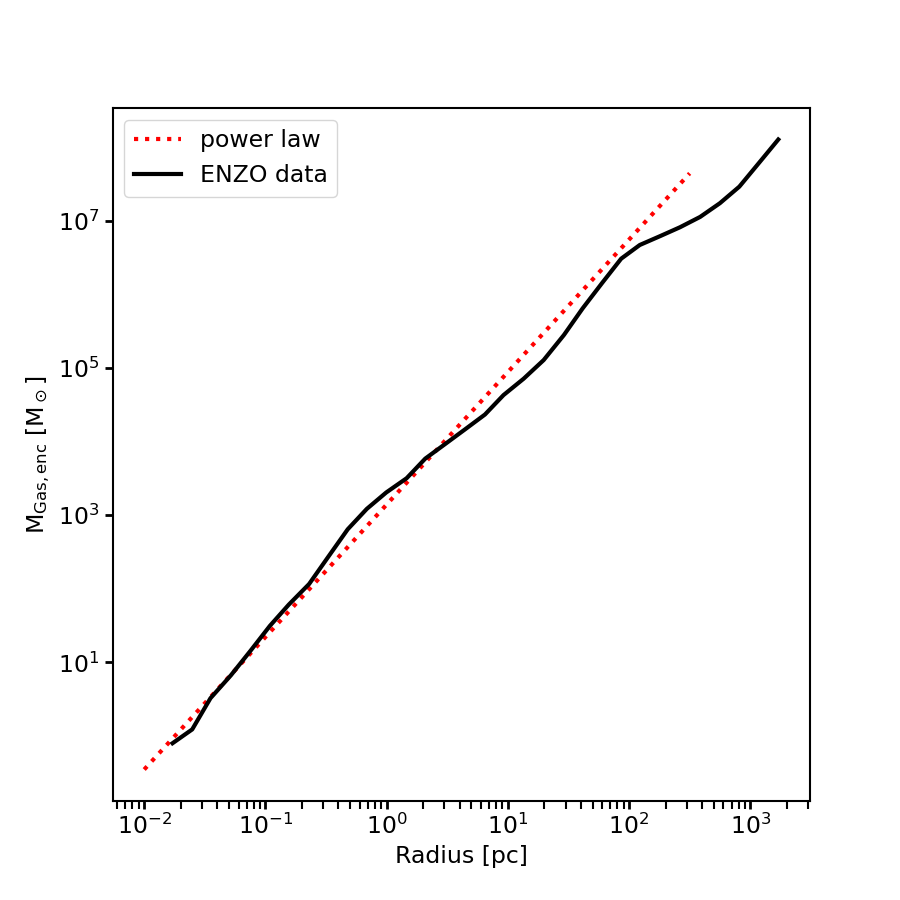}
      \caption{Radial profile of the enclosed gas mass at the end of the \textsc{ENZO} run (black solid line),
        along with the enclosed mass for the power-law potential given by Eq.(\ref{eq:power_law_pot}) (dotted red line), which is used as a background potential to mimic the gas in our models.}
         \label{fig:ext_pot}
   \end{figure}

To model the gravitational effects of the gas on the cluster, we employ an analytic background potential of the form described in \cite{BinneyTremaine2008}:
   
   \begin{equation}
       \label{eq:power_law_pot}
       \rho(r) = \rho_0 \left( \frac{r_0}{r}\right)^{\alpha},
   \end{equation}
and the enclosed mass given as:
   \begin{equation}
       \label{eq:bkg_pot_enclosed_mass}
       M(<r) = \frac{4\pi \rho_0 r_0^{\ \alpha} r^{\ 3-\alpha}}{3-\alpha}.
   \end{equation}
We set $\rho_0=1105.243$~M$_\odot$~pc$^{-3}$, $r_0=0.6$~pc, and $\alpha=1.1$, to match the enclosed mass at 2~pc from the \textsc{ENZO} run, as shown in Fig.~\ref{fig:ext_pot}.

 We further mimic gas accretion onto the cluster by modifying $\rho_0$ as follows:
   \begin{equation}
       \rho_0(t) = 1105.243 \ {\rm M_\odot \ pc^{-3} } + \dot{\rho}_0 t,  
   \end{equation}
    and
   \begin{equation}
       \dot{\rho}_0 = \dot{M}_{\rm gas} \frac{3-\alpha}{4\pi r_0^{\ \alpha} r^{\ 3-\alpha} },
   \end{equation}
with $r=2$~pc and $\dot{M}_{\rm gas}=2$~M$_\odot$~yr$^{-1}$ taken from the \textsc{ENZO} run. This results in a value of $\dot{\rho}_0=0.142$~M$_\odot$~pc$^{-3}$~yr$^{-1}$, which is used as the inflow rate for five models. We also explore a case with a larger inflow rate of  $\dot{M}_{\rm gas}=0.9$~M$_\odot$~yr$^{-1}$ at $r=1$~pc, resulting in $\dot{\rho}_0=0.238$~M$_\odot$~pc$^{-3}$~yr$^{-1}$.

\subsection{The IMF}
\begin{figure}[h]
   \centering
   \includegraphics[width=9cm]{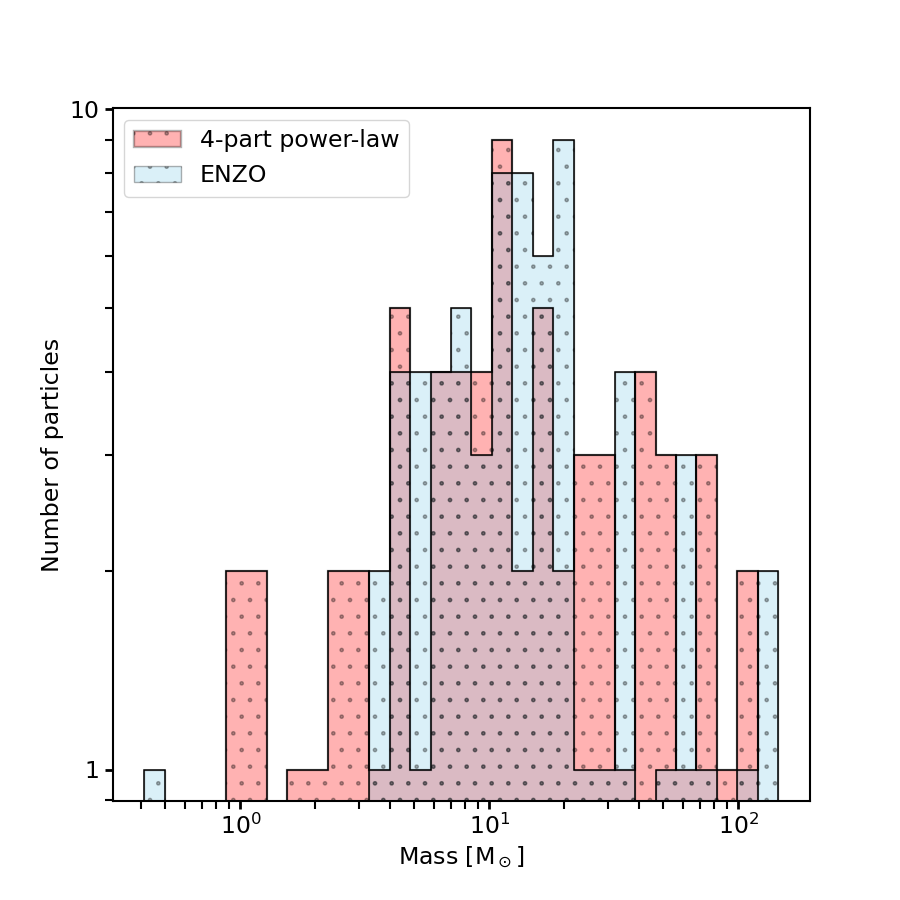}
      \caption{Comparison of the IMF at the end of the \textsc{ENZO} run, with a sampling of the 4-part power law IMF defined in Eq.(\ref{eq:3part_pl_imf}). Both samples contain total of 1497~M$_\odot$ distributed among 68 particles. The 4-part power law IMF is used to draw stars during star formation episodes in our simulations.}
         \label{fig:compare_imfs}
   \end{figure}

We construct an IMF  based on the output of the \textsc{ENZO} run and use it for all models. For this purpose, we employ a (4-part) power law of the form:
   \begin{equation}
        \label{eq:3part_pl_imf}
        \xi (m) \propto m^{-\alpha_i},
    \end{equation}
    where
    \begin{align*} 
        \alpha_1 &=  \ \ \ 0.66, & 0.4\leq & \ \ m/{\rm M_\odot} < 3.0, \\ 
        \alpha_2 &=  -0.15, & 3.0\leq & \ \ m/{\rm M_\odot}  < 10.0, \\
        \alpha_3 &=  \ \ \ 0.83, & 10.0\leq & \ \ m/{\rm M_\odot}  < 20.0,\\
        \alpha_4 &=  \ \ \ 1.75, & 20.0\leq & \ \ m/{\rm M_\odot}  < 150.0.  \end{align*} 
We present the comparison of constructed power law IMF with the \textsc{ENZO} output in Fig.~\ref{fig:compare_imfs}.

To account for unresolved or subsequent star formation (SF), we assume a fixed star formation efficiency ($\epsilon_{\rm SF}$) for 2 Myr. This determines the total stellar mass for our models as  M$_{\rm IMF} = \epsilon_{\rm SF} \times \dot{M}_{\rm gas}\times2$~Myr. By varying $\epsilon_{\rm SF}$, we obtain three different values for M$_{\rm IMF}$. The adopted values of$\epsilon_{\rm SF}$ and the resulting M$_{\rm IMF}$ are listed in Table~\ref{tab:sim_set}.


   \subsection{Star formation}
    For five of our models, we treat star formation as a continuous process lasting for $t_{\rm SF}=2$~Myr, which we refer to as continuous SF models. New stars are inserted into the simulation at time-intervals of
   \begin{equation}
       \label{eq:dt_SF}
       dt_{\rm SF}= \frac{M_{\rm SF}}{\epsilon_{\rm SF} \dot{M}_{\rm gas}},
   \end{equation}
   with $M_{\rm SF}=200$~M$_\odot$ being the total mass in newly formed stars. 
   The value of M$_{\rm SF}$ is set to be higher than the maximum mass in the adopted IMF, ensuring that high-mass stars can also be created.
   For $\epsilon_{\rm SF}=0.1$, $dt_{\rm SF}=1000$~yr. 
   We also include a model for which the SF process lasts for only $t_{\rm SF}=0.5$~Myr, which we refer to as the rapid SF model. We achieve this by fixing $dt_{\rm SF}=1000$~yr, and allowing a maximum mass of $M_{\rm SF}=800$~M$_\odot$ per SF episode, such that the same total number of stars are formed but in a shorter time span.

   During the simulations we define the star formation efficiency as
   \begin{equation}
       \epsilon_{\rm SF \ now} = \frac{M_{\rm stars}}{M_{\rm gas,2} + M_{\rm stars}},
   \end{equation}
   where $M_{\rm gas,2}$ is the gas mass enclosed within the central 2~pc, and $M_{\rm stars}$ is the current stellar mass. The amount of gas that is converted into new stars $M_{\rm new \ stars}$ is defined as:

   \begin{eqnarray}
      M_{\rm new \ stars} = \left\{  \begin{aligned} & \ \epsilon_{\rm SF}M_{\rm gas,2} - M_{\rm stars}(1-\epsilon_{\rm SF}), & \epsilon_{\rm SF \ now}<\epsilon_{\rm SF} ,  \\
      & \  \epsilon_{\rm SF}\dot{M}_{\rm gas}dt_{\rm SF},  & \epsilon_{\rm SF \ now} \geq \epsilon_{\rm SF},  \end{aligned} \right.
   \end{eqnarray}   
   where $\epsilon_{\rm SF}$ is the SF efficiency assumed for our model, as presented in Table~\ref{tab:sim_set}. The use of $\epsilon_{\rm SF \ now}$ allows us to maintain a SF efficiency close to $\epsilon_{\rm SF}$ during the entire simulation.

   Once the total mass in new stars has been decided, we draw stars from the IMF until the required mass has been reached. A safe check is imposed such that if $M_{\rm new \ stars}>M_{\rm gas,2}$, then star formation is postponed to the next timestep.

The newly formed stars are initialized with a Plummer density profile \citep{Plummer1911} with a  $R_{\rm v}=1.0$~pc (scale length $a=0.589$~pc) and in virial equilibrium, considering all the mass enclosed within 1~pc. We also impose a cut-off radius of $r_{\rm cut}=3.5a=2.06$~pc to prevent new stars from forming too far from the cluster center. The mass transformed into stars is removed from the background potential by modifying $\rho_0$ as follows:
   \begin{equation}
       \label{sec:take_mass_from_bkg_pot}
       \rho_{0,f} = \rho_{0,i} - \frac{(3-\alpha)M_{\rm new \ stars}}{4\pi r_0^{\ \alpha} r^{\ 3-\alpha} },
   \end{equation}
   with $r=2$~pc.
   
\subsection{Gas accretion by individual stars}
We model gas accretion onto the individual stars by assuming the Bondi-Hoyle-Littleton accretion rate. For this, we first calculate the Bondi radius as follows:
   \begin{equation}
       \label{eq:bondi_radius}
       r_{\rm bondi} = \frac{2Gm}{c_s^2 + v^2},
   \end{equation}
   where $m$ is the mass of the accreting star, $v$ is the velocity of the star, and $c_s$ is the sound speed. For the sound speed, we assume constant values of $\gamma=5/3$, $T=200$~K, and $\mu=1.43$.
   The Bondi radius is used to calculate the accretion rate as :
   \begin{equation}
      \label{eq:bondi_mod}
       \dot{m} = \frac{4 \pi G^2 m^2 \rho_0}{(v^2 + c_s^2)^{3/2}} \left(\frac{r_0}{r+r_{\rm bondi}} \right)^{\alpha},
   \end{equation}
 where $r$ is the radial position of the accreting star. The use of the Bondi radius prevents a singularity at $r=0$. The accretion rate is recalculated at each timestep and an upper limit of $\dot{m}_{\rm max}=10^{-4}$~M$_\odot$~yr$^{-1}$ is imposed for main sequence stars. We also consider one model in which we set $\dot{m}_{\rm max}=10^{-3}$~M$_\odot$~yr$^{-1}$. The accreted mass is removed from the background potential by modifying $\rho_0$.\\

To prevent dramatic changes in individual stellar masses, we use an adaptive timestep approach that imposes a maximum mass increase of $10$~M$_\odot$ per timestep. At the end of each timestep, the required timestep is computed as follows: 
\begin{equation}
       dt_{\rm req} = \frac{10\ {\rm M_\odot}}{{\rm max}(\dot{m}_i)},
\end{equation}
where max($\dot{m}_i$) is the maximum accretion rate among all the stars. If $dt_{\rm req} \leq dt$, the current timestep is modified as
\begin{equation}
      dt_{\rm new} = dt \cdot 2^{-j},
\end{equation}
where $j$ is the minimum integer for which the condition $dt_{\rm new}<dt_{\rm req}$ is fulfilled. On the other hand, if $dt_{\rm req}\geq 2dt$, then $dt_{\rm new}=2dt$.
We note that this timestep reduction is rarely triggered in our simulations and is only relevant for high-mass stars.

\subsection{Stellar collisions and the associated mass loss}
\label{sec:collisions_mass_loss}
Stellar collisions are detected during the integration with the $N$-body code \textsc{PH4}. A stellar collision occurs if 
\begin{equation}
     d\leq R_1 + R_2,
\end{equation}
where $d$ is the separation between two stars, and $R_1$ and $R_2$ are the stellar radii. Once a stellar collision is detected we replace the colliding particles by a new particle placed at the center of mass, with a velocity calculated assuming linear momentum conservation. We include a mass loss recipe based on the work of \citet{Glebbeek2013}. The mass loss depends on the mass ratio $q=m_1/m_2$ (with $m_2\geq m_1$), and on the stellar structure through the parameters $R^{0.86}$ and $R^{0.5}$, which represent the radii containing 86\% and 50\% of the stellar mass, respectively. We assume that all merging stars are on the main sequence and model them as $n=3$ polytropes to obtain the required parameters. Using this approach, the fraction of mass lost during a collision can be calculated as follows:
   \begin{eqnarray}
      \label{eq:mass_loss_frac_adopt}
      f_{\rm lost} = \frac{q}{1+q^2} \left\{  \begin{aligned} & \ 0.243, & q<0.4 ,  \\
      & \  0.3,  & q\geq 0.4.  \end{aligned} \right.
   \end{eqnarray}
   The mass of the collision product is estimated as
   \begin{equation}
       m_{\rm new} = (1-f_{\rm lost}) (m_1 + m_2).
   \end{equation}
The mass lost during the collision is also assumed to be lost from the system. The stellar radius of newly formed star is calculated using the mass radius relation in Sec.~\ref{sec:MR_relation}, with the accretion rate taken as the maximum of accretion rate among the colliding stars. 

   \begin{table*}
      \caption[]{List of all the models with the corresponding parameters. The first column is the model ID, the second column shows the total time during which SF occurs, the third column is the assumed SF efficiency, the fourth column is the density increase of the background potential due to gas accretion onto the cluster, the fifth column is the total mass of the IMF, the sixth column is the maximum accretion rate allowed for individual stars, the seventh column is the radius of the cluster, the eight column is the number of stars in the IMF, the ninth column is the average mass of the most massive object, the tenth column is the average number of collisions, and the eleventh column is the average amount of total mass lost by the MMO during stellar collisions. Three simulations were performed for each model. }
         \label{tab:sim_set}
         \centering          
         \begin{tabular}{l c c c r c c r r r r}     
         \hline\hline       
         Model & $t_{\rm SF}$ & $\epsilon_{\rm SF}$ & $\dot{\rho}_0$ & $M_{\rm IMF}$\ \ \ \ \ & $\dot{m}_{\rm max}$ & $R_{\rm v}$ & $N_{\rm stars}$ & $\overline{\rm M}_{\rm MMO}$ & $\overline{\rm N}_{\rm col}$ & $\overline{\rm M}_{\rm lost, coll}$\\
             & Myr           &                     & M$_\odot$~yr$^{-1}$~pc$^{-3}$ & M$_\odot$\ \ \ \ \ \ & M$_\odot$~yr$^{-1}$ & pc &  & M$_\odot$ & &M$_\odot$ \\
         \hline                    
         E1 & 2.0 & 0.1 & 0.142 & $400\,610.8$ & $10^{-4}$ & 1.0 & $7100$ & $440 \pm 21$ & $5 \pm 1$ & $70 \pm 5$\\   
         E2 & 2.0 & 0.2 & 0.142 & $800\,929.6$ & $10^{-4}$ & 1.0 & $31\,500$ & $756 \pm 215$ & $19 \pm 6$ & $226 \pm 103$ \\
         E3 & 2.0 & 0.3 & 0.142 & $1\,203\,113.3$ & $10^{-4}$ & 1.0 & $47\,500$ & $2266 \pm 1710$ & $42 \pm 12$ & $876 \pm 741$\\
         E4 & 2.0 & 0.1 & 0.142 & $400\,610.8$ & $10^{-3}$ & 1.0 & $7100$ & $14434 \pm 3098$ & $30 \pm 7$ & $5323 \pm 1310$\\
         E5 & 2.0 & 0.1 & 0.238 & $400\,610.8$ & $10^{-4}$ & 1.0 & $7100$ & $332 \pm 12$ & $3 \pm 1$ & $0.0 \pm 0.0$\\
         E6 & 0.5 & 0.1 & 0.142 & $400\,610.8$ & $10^{-4}$ & 1.0 & $7100$ & $4704 \pm 21$ & $41 \pm 3$ & $1766 \pm 25$\\
         \hline                  
      \end{tabular}
   \end{table*}

\section{Results}
\label{sec:results}

We present here the main findings of our models, beginning with a description of the evolution of the star clusters and the formation of the most massive objects (MMOs). We then discuss the processes that lead to variations in the final masses of the MMOs.

\subsection{Cluster and core contraction}
\begin{figure}
\includegraphics[width=9cm]{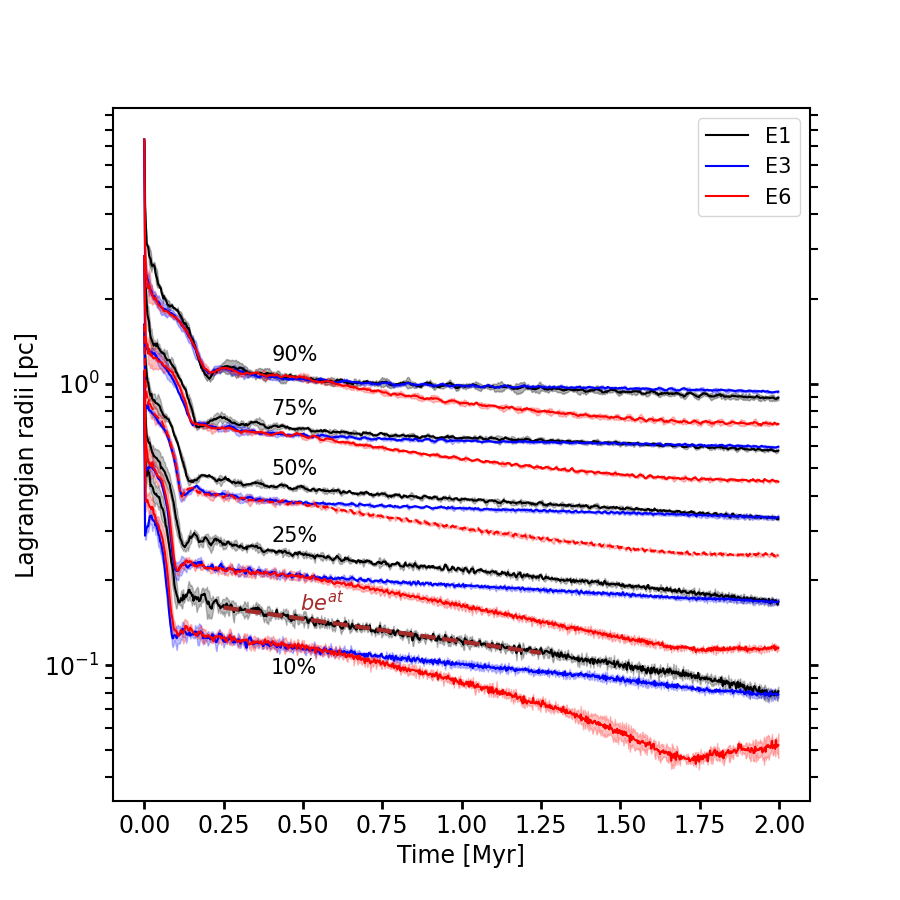}
    \caption{Lagrangian radii evolution of some star cluster models. Solid lines represent the average of three simulations and the shaded area represents one sigma errors. A fit to the core radius of the form $R_c=b e^{at}$ is shown with a brown dashed line. We find that the core contraction rate depends on the star formation efficiency. Additionally we see the signature of a core collapse for models in which all stars form in a shorter time span (E6).}
    \label{fig:contraction_clusters}
\end{figure}

Our simulations show a continuous contraction of the cluster core in all models. However, we note that the contraction rate depends on the $\epsilon_{\rm SF}$ and the timescale over which star formation occurs. In Fig.~\ref{fig:contraction_clusters}, we present the Lagrangian radii of our simulated clusters as a function of time for different models. Each line represents the average of three simulations, with the shaded region indicating one-sigma errors. A clear difference is immediately apparent in model E6 with rapid SF, where the clusters undergo core collapse, as traced by the 10\% Lagrangian radius. For the remaining models, the clusters contract globally and continuously without reaching core collapse.

We also compare the contraction rates of the clusters by fitting an exponential function of the following form:
\begin{equation}
    R_{c} = b\ \exp{(a t)},
\end{equation}
to the 10\% Lagrangian radii, for all the clusters. Here $a$ and $b$ are constants to be determined by the fit. The fit is performed between 0.25 and 1.25 Myr for all clusters, except for the models with rapid SF, where we select the interval 0.25–0.6 Myr to avoid the core collapse phase, as illustrated in Fig.~\ref{fig:contraction_clusters}. The parameter $a$ indicates the contraction rate of the cluster. We find that the contraction rate is anti-correlated with $\epsilon_{\rm SF}$, i.e., meaning faster contraction occurs for lower $\epsilon_{\rm SF}$. This result can be understood in terms of the relaxation timescale of a star cluster, which is given by \citep{Spitzer1987}
\begin{equation}
    t_{\rm rh} = 0.138 \frac{N}{\ln({\gamma N})} t_{\rm cross},
\end{equation}
with $N$ being the number of stars and $t_{\rm cross}$ the crossing time. We therefore expect $t_{\rm rh}$ to be longer in a cluster with more stars (provided $t_{\rm cross}$ remains unchanged). Thus, when $\epsilon_{\rm SF}$ is higher, $N$ is larger and $t_{\rm rh}$ is longer. We also find that the contraction rate between 0.25-0.6 Myr  is slower in clusters with rapid SF. The same argument applies here; the number of stars during the time span considered for this analysis is higher, which increases the relaxation time. As a result, the cluster contracts more slowly.

For the case of a higher gas inflow rate into the cluster, i.e., model E5 ($\dot{\rho}_0=0.238$~M$_\odot$~yr$^{-1}$~pc$^{-3}$), we also see a slower contraction rate for the cluster core. Two factors contribute to this effect. On one hand, the gas reservoir grows more rapidly, leading to more frequent episodes of star formation, as implied by Eq.(\ref{eq:dt_SF}). This results in a larger $N$ at any given time, thus increasing the relaxation timescale and slowing the core contraction.
On the other hand, a higher gas inflow rate leads to more massive clusters, where stellar velocities are higher compared to those in clusters with lower inflow rates \citep{Kroupa2020}. Higher stellar velocities delay the relaxation processes \citep{Reinoso20}, further slowing the core contraction rate.

\subsection{Core collapse of clusters with rapid star formation}
\begin{figure}
 \centering
 \includegraphics[width=\hsize]{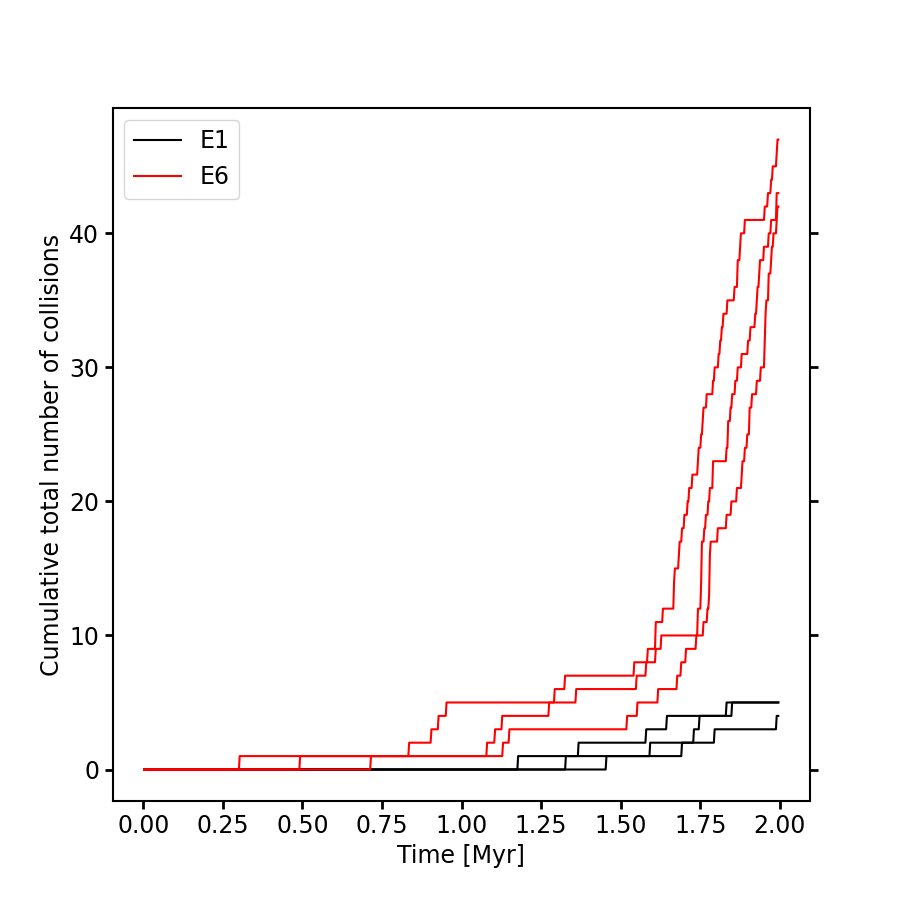}
    \caption{Cumulative number of collisions as function of time for two models that differ only in the timescale over which star formation occurs. We see the signature of runaway stellar collisions in clusters for which star formation occurs in a sorter time span.}
       \label{fig:ncol_ncore}
\end{figure}

From Fig.~\ref{fig:contraction_clusters} it is clear that for clusters with rapid SF, a core collapse occurs at around 1.75~Myr. This effect is also evident in the average mass of the MMO in Table~\ref{tab:sim_set}. The only difference between models E1 and E6 is the timescale over which SF occurs, yet there is a factor $\sim10$ difference in the mass of the MMO. This difference arises from core collapse, which increases stellar density and, consequently, the number of collisions, as shown in Fig.~\ref{fig:ncol_ncore}.
We note that the cores of clusters with rapid star formation tend to be more compact even before the onset of core collapse, which increases the collision probability, as reflected in Fig.~\ref{fig:ncol_ncore}. Additionally, the first collisions tend to occur between 0.25 and 0.75~Myr for clusters with rapid star formation, while they tend to occur later, between 1.25 and 1.5~Myr, for clusters with continuous star formation. The high collision rate seen at $\sim1.75$~Myr in Fig.~\ref{fig:ncol_ncore} is also a distinctive feature of the core collapse in star clusters that triggers a runaway collision process \citep[see e.g.][]{PortegiesZwart2004,Sakurai2017,Reinoso20,Vergara23}.

\begin{figure}
\includegraphics[width=\hsize]{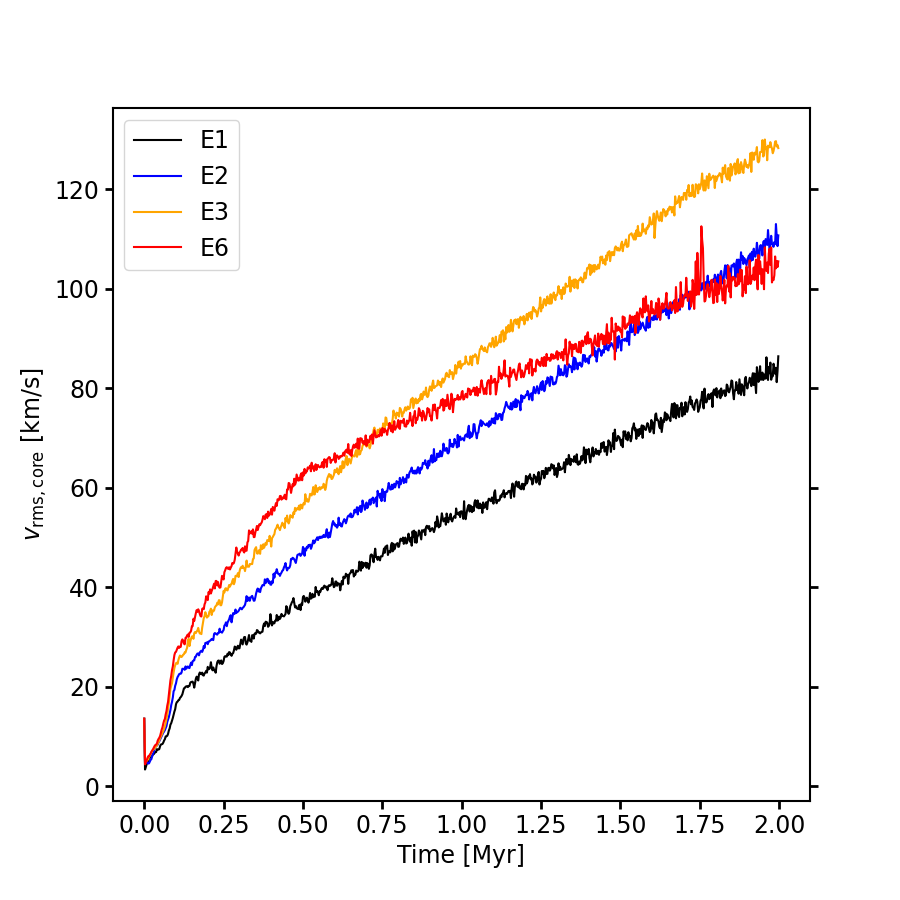}
    \caption{Root-mean-square velocity of stars in the cluster core as function of time for all the models with $\dot{\rho}_0=0.142$~M$_\odot$~yr$^{-1}$~pc$^{-3}$. The rate for the increase in v$_{\rm rms, core}$ correlates with the star formation efficiency. We also note that after star formation ceases at 0.5~Myr for model E6 (rapid star formation), the rate at which v$_{\rm rms, core}$ increases is slower.}
    \label{fig:vrms_M_clusters}
\end{figure}

Now, we explain why a core collapse occurs when star formation is rapid. It is important to recall that in our model, stars are initialized with a Plummer density profile with $R_{\rm v}=$1~pc, and in virial equilibrium, accounting for all the mass enclosed within the central parsec.  From the virial theorem $\sigma_v \propto M$, meaning that the larger the mass enclosed, the higher the velocity dispersion of the newly formed stars. This implies that for higher $\epsilon_{\rm SF}$, $\sigma_v$ increases more rapidly, as shown in Fig.\ref{fig:vrms_M_clusters}. For clusters with rapid star formation, the  $v_{\rm rms}$ of stars in the core no longer increases as quickly once star formation ceases. We interpret this result as follows: the continuous injection of stars in virial equilibrium effectively provides a source of kinetic energy into the cluster. Once this source is removed, the velocity changes of the stars are driven solely by the growing background potential and two-body relaxation effects with existing stars. Without this continuous kinetic energy injection, the clusters are able to undergo relaxation, which can lead to core collapse over a shorter timescale.


\subsection{Emergence of the most massive object}
\begin{figure*}
\includegraphics[width=\hsize]{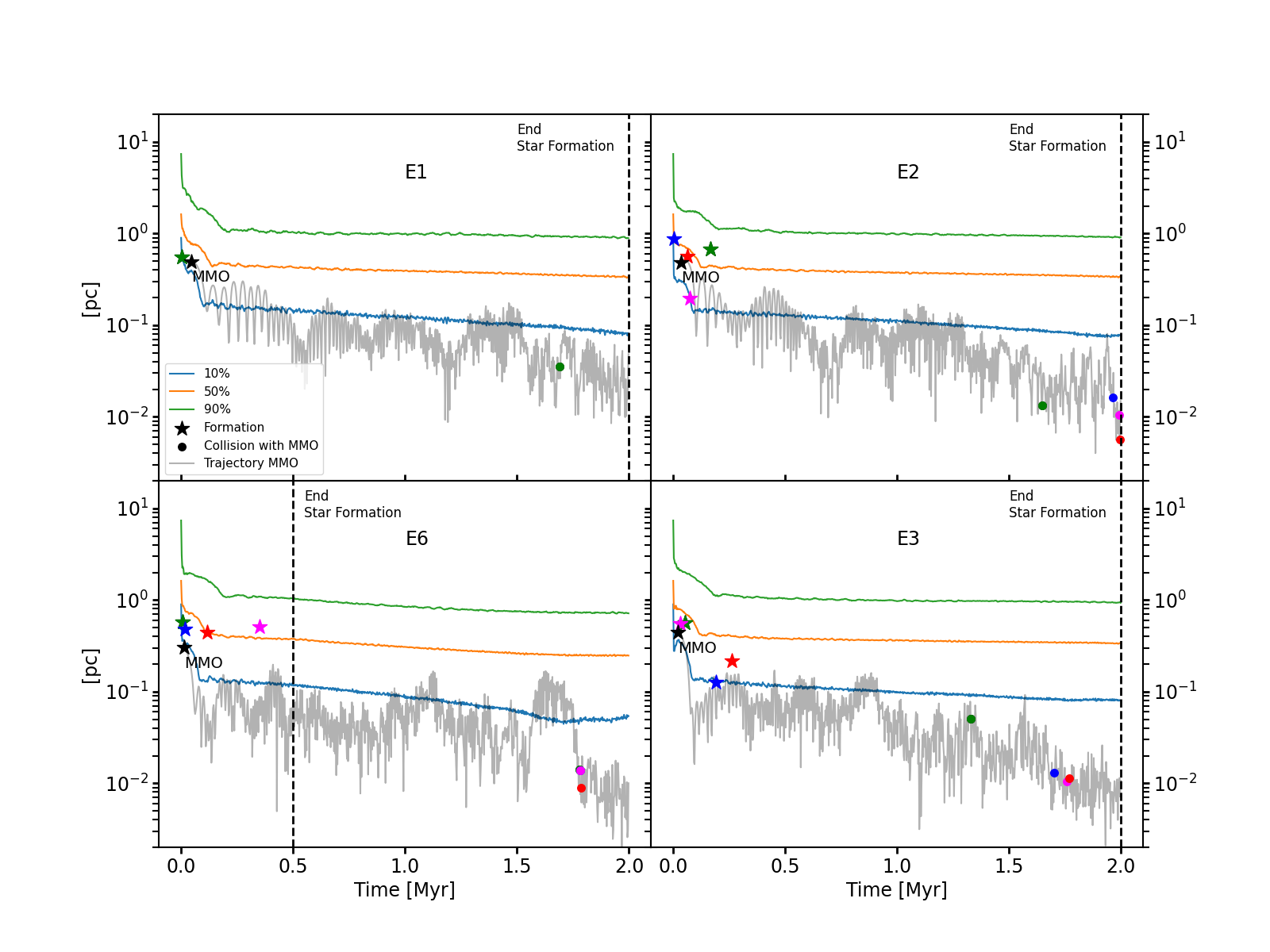}
    \caption{ Lagrangian radii evolution (10\%,50\%,90\%), radial position of the MMO (gray line), radial position at formation for stars that collide with the MMO (star symbols), and radial position of the same stars a the moment of the collision (filled circles); for simulations with different $\epsilon_{\rm SF}$ and for a simulation with rapid SF (E6). In general we see that the object that becomes the MMO is formed early on in the cluster evolution and rapidly sinks to the cluster core, where eventually experiences the collisions that turn it into the MMO. The formation of the MMO is marked with a black star symbol. The moment at which SF ends is indicated by a dashed vertical black line.}
    \label{fig:mmo_formation}
\end{figure*}

\begin{figure*}
\includegraphics[width=\hsize]{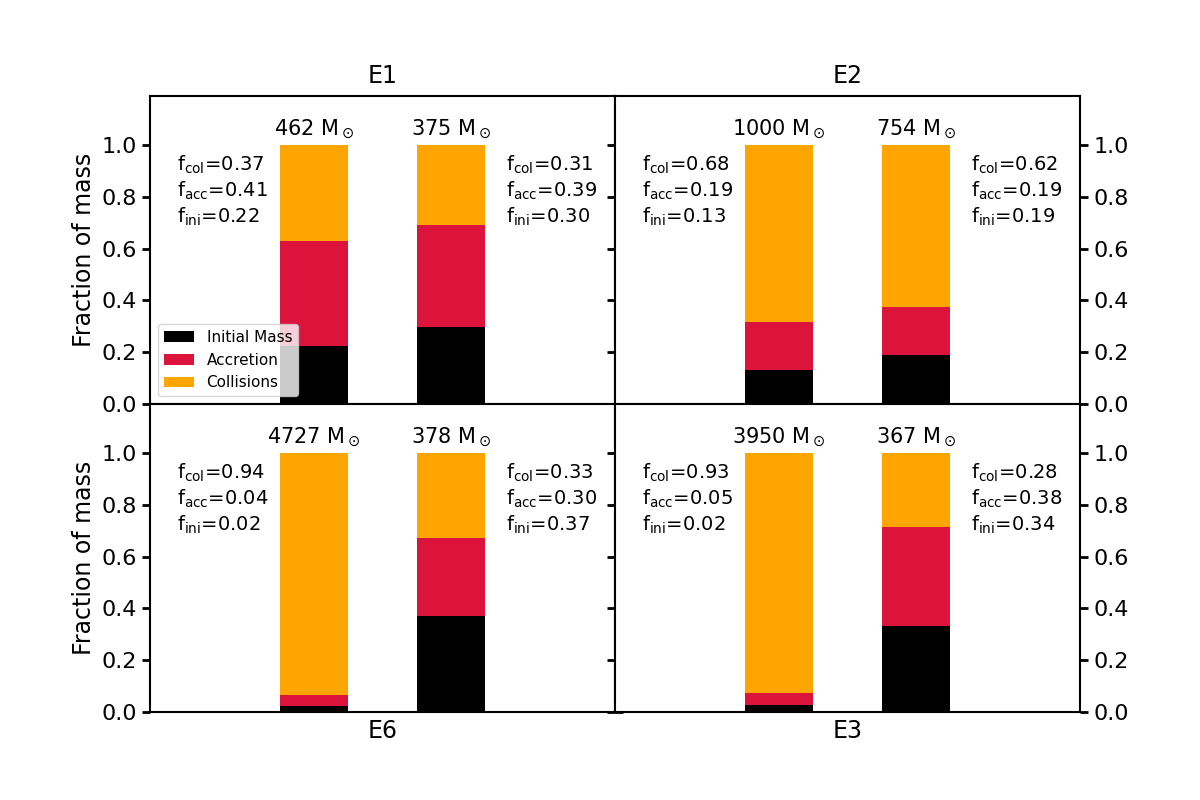}
    \caption{Fraction of the final mass gained through collisions $f_{\rm col}$, accretion $f_{\rm acc}$, and the initial mass $f_{\rm ini}$, for the MMO and second MMO, for models with different $\epsilon_{\rm SF}$. We present one simulation in each panel. In general the initial mass is always in the range 80-150~M$_\odot$, and the mass gained by accretion is always in the range $100-200$~M$_\odot$, the difference comes from the total mass gained through collisions.}
    \label{fig:mmo_mass_fracs}
\end{figure*}

In all our simulations, we find that the object that becomes the MMO forms early, within the first 50 kyr of evolution, and is formed outside the cluster core but within the half-mass radius. This object initially has a mass ranging from 80-150~M$_\odot$, i.e., in the high-mass end of the IMF, and gradually sinks to the cluster core. An example of this process for different simulations is presented in Fig.~\ref{fig:mmo_formation}. For $\dot{m}_{\rm max}=10^{-4}$~M$_\odot$~yr$^{-1}$, the MMO consistently gains a total of 100-200~M$_\odot$ 
through gas accretion, while the remainder of its mass growth comes from collisions with other stars as shown in Fig.~\ref{fig:mmo_mass_fracs}. The relative contribution from accretion (collisions) is lower (higher) for more massive objects. Collisions with the MMO tend to occur rather late in the simulation at $t>1.25$~Myr.

Our simulations show that most of the stars in the high-mass end of the IMF, i.e.,with an initial mass $M>100$~M$_\odot$, typically form outside the core and therefore never experience a collision. In fact, the fraction of these stars that survive until the end of the simulation is typically $>98$\%. Moreover, stars that experience a collision are typically the ones that are formed in between the cluster core and the half-mass radius, whereas the majority of the stars formed in the core ($\gtrsim90$\%) survive until the end of the simulation. This means that even though the collisions occur in the cluster core, the colliding stars are typically formed outside. This is somewhat different in model E6, in which the clusters experience core collapse. In this case, the survival rate of $M>100$~M$_\odot$ stars is $\sim95\%$, and the majority of the high mass stars in the core ($>60$\%) experience a collision.

\begin{figure}
\includegraphics[width=\hsize]{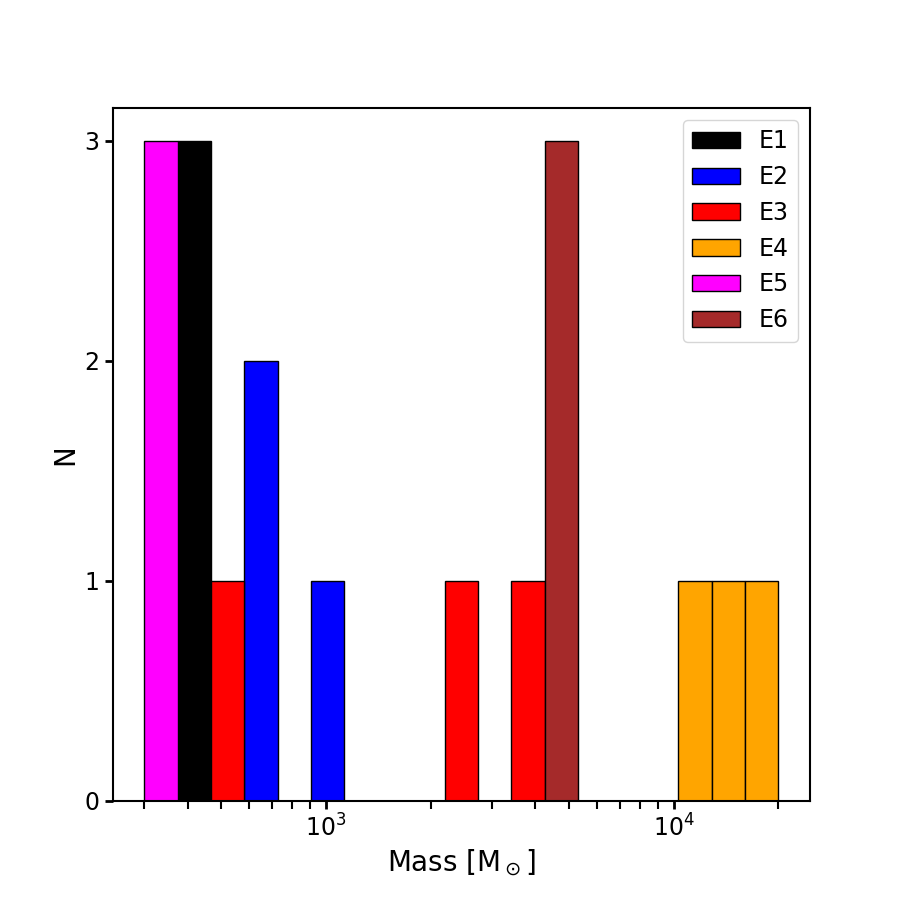}
    \caption{Mass distribution of the MMOs formed in each of our models, including all 3 simulations per model. In general the mass of the MMO increases with the star formation efficiency, as well as for higher accretion rates onto stars (E4), or when forming all the stars in a shorter timespan (E6).}
    \label{fig:final_mmo_mass_distr}
\end{figure}

The stochastic nature of stellar collisions makes it difficult to identify a single factor that can predict which object will become the MMO. Moreover, it is always a collision with a star of similar mass that marks the emergence of the MMO in the core. As shown in Fig. \ref{fig:final_mmo_mass_distr}, we find that our star cluster models produce stars with masses greater than $10^3$~M$_\odot$ when $\epsilon_{\rm SF} \geq 0.3$, or when a burst of star formation assembles the cluster in $\leq 0.5$ Myr. If high accretion rates onto individual stars are sustained (i.e., E4 model), the MMO can reach masses $\geq 10^4$~M$_\odot$.

\subsection{Higher accretion rate onto stars}
Here, we present the case of a higher accretion rate onto the stars. Model E4 is identical to E1, except for the maximum accretion rate of $\dot{m}_{\rm max}=10^{-3}$~M$_\odot$~yr$^{-1}$ in the former case. 
 We note that the final mass of the MMO is approximately 30–40 times larger in E4. This is due to both a larger mass gained through accretion (by a factor of $\sim5$) and a greater number of collisions (by a factor of $\sim8$). Additionally, the typical mass of stars colliding with the MMO is around 1000~M$_\odot$ due to the higher accretion rate.


The emergence of the MMO is similar to other models. A star with an initial mass of $70-120$~M$_\odot$ forms between the cluster core and the half-mass radius, and then sinks to the center as it gains mass through accretion. The key difference in this case is that the mass growth occurs much more rapidly, with the star typically gaining $\sim 1400$~M$_\odot$ by accretion. Stellar collisions begin at earlier times, typically around 0.5–0.7 Myr, similar to the model with rapid SF. Notably, our results suggest that the onset of stellar collisions occurs once the number density of stars reaches $10^5$~pc$^{-3}$, with this quantity being a more reliable predictor than the stellar mass density.


\subsection{MMOs in binary systems}
We investigate whether the MMOs are part of a binary system at the end of the simulations. To do this, we first identify all the stars that are gravitationally bound to the MMO in the final snapshot. A pair of stars is considered gravitationally bound if their specific orbital energy ($\epsilon$) is negative, as given by:
\begin{equation}
    \label{eq:specific_orbital_energy_2}
    \epsilon = \frac{1}{2} v^2 - \frac{\mu}{r},
\end{equation}
where $v$ is the relative velocity, $r$ is the distance between the pair,  and $\mu=G(m_1+m_2)$.
We found that in most of our models, multiple stars are bound to the MMO. However, we are primarily interested in long-lived binary systems, not just gravitationally bound pairs. Therefore, we also consider the perturbations induced by the other stars as they can disrupt the binary system. To this end, we calculate a perturbation parameter, which we define as:
\begin{equation}
    \gamma = \frac{ {\rm max}(F_{\rm p,max} , F_{\rm s,max}) }{F_{p,s}},
\end{equation}
where $F_{\rm p,max}$ is the maximum force acting on the primary star (i.e. the most massive one) due to the rest of the stars (excluding the secondary). $F_{\rm s,max}$ is analogous to $F_{\rm p,max}$ but for the secondary star. $F_{p,s}$ is the gravitational force between the primary and secondary star. 

To classify a bound pair as an unperturbed binary, we impose the condition $\gamma<0.1$, which ensures that we are considering pairs of particles that are dynamically separated from the rest of the system. We note that this perturbation parameter has been previously used to determine when to apply regularization in the \textsc{NBODY} code series \citep{Nbody_Sverre,Superzem_Nbody6pp,Wang_Nbody6ppgpu,Nitadori2012,Kamlah22}. Another criterion for selecting potentially long-lived binaries is to compare the binary's semi-major axis to the hard-soft boundary $a_{\rm HS}$, which we compute as:
\begin{equation}
    a_{\rm HS} = \frac{2GM}{v_{\rm vrms,01}^2},
\end{equation}
with $M$ being the mass of the MMO, and $v_{\rm vrms,01}$ the root-mean-square velocity of all the stars in a sphere of $r=0.1$~pc centered on the MMO. In general, hard binaries ($a<a_{\rm HS}$) tend to harden, while soft binaries ($a>a_{\rm HS}$) tend to soften \citep{Heggie75,Hills75}. We therefore also impose the condition $a/a_{\rm HS}<1$ in order to classify the pair as a potentially long-lived binary.

In Table~\ref{tab:binary_mmo}, we list the number of stars bound to the MMO, the minimum value of $\gamma$ among the bound stars ($\gamma_{\rm min}$), the mass of the star for which $\gamma=\gamma_{\rm min}$, and the ratio $a/a_{\rm HS}$. Based on the adopted criteria, we identify only three potentially long-lived binaries in our simulations.

   \begin{table*}
      \caption[]{Parameters related to the identification of binary systems that involve the MMO in each simulation. Column 1 presents the corresponding simulation, column 2 indicates the number of stars that are gravitationally bound to the MMO, column 3 shows the minimum value of the perturbation parameter $\gamma_{\rm min}$ among the bound stars, column 4 indicates the mass of the star for which $\gamma=\gamma_{\rm min}$, column 5 shows the mass of the MMO, and column 6 presents the ratio of the binary semi-major axis to the semi-major axis at the hard-soft boundary.}
         \label{tab:binary_mmo}
         \centering          
         \begin{tabular}{l r l r r r}
         \hline \hline
            Simulation & N$_{\rm bound}$ & $\gamma_{\rm min}$ & M ($\gamma_{\rm min}$)  & M$_{\rm MMO}$ & $a/a_{\rm HS}$\\
                                 &                 &                    &     M$_\odot$ &  M$_\odot$  & \\
         \hline
           E1\_1     & 0              &  -              & -      & 462 & -\\
           E1\_2     & 2              &  1.88           & 232 & 439 & 18.28\\
           E1\_3     & 2              &  7.53           & 292 & 420 & 66.68\\
                      &                &                 &         &           & \\ 
           E2\_1     & 8              &  0.037          & 754 & 1000 & 0.80\\
           E2\_2     & 6              &  0.83           & 329 & 591 & 4.60\\
           E2\_3     & 3              &  44.50          &  98 & 681 & 27.17\\
                      &                &                 &         &           & \\ 
           E3\_1     & 43             &  1.19           & 226 & 3950 & 1.64\\
           E3\_2     & 1              &  2255           &  29 & 532 & 366.99\\
           E3\_3     & 19             &  0.116          & 251  & 2317 & 0.31 \\
                      &                &                 &         &           &     \\ 
           E6\_1     & 34             &  0.009          & 335 & 4727 & 0.09 \\
           E6\_2     & 30             &  0.003          & 231 & 4699 & 0.03\\ 
           E6\_3     & 40             &  0.92           & 231 & 4685 & 0.57 \\
          \hline

    \end{tabular}
   \end{table*}

\section{Conclusion and Discussion}
\label{sec:Discussion_and_conclusions}
Our models of Pop III star clusters are based on realistic initial conditions taken directly from a cosmological simulation. We have developed a model for their subsequent assembly and evolution. We find that BHs of up to a few thousand solar masses are formed through this scenario. These BH seeds are heavier  compared to the single Pop III remnant scenario. Additionally, the fact that the BH is embedded in a massive star cluster will further aid its sinking into the deep potential well of the galaxy \citep{Mukherjee24}. In particular, we might expect the system of stars and BH to experience subsequent episodes of gas accretion. BHs in such environments have been shown to grow at rates that may exceed the Eddington limit, with e-folding times of 200-300 Myr \citep{Partmann25}. This pathway, therefore, offers a promising alternative to the DCBH scenario.  The higher abundance of BH seeds forming through this channel can explain the recently observed abundance of AGN at high $z$ as found in high-resolution cosmological simulations \citep{Mayer25}, along with the potential for further efficient growth. However, certain effects are still not included in our models, and we discuss some of these caveats below.

We have neglected mass loss due to stellar winds, given the zero metallicity of the stars considered in our models \citep{Krtivka2006}. It is worth noting that although rotating Pop III star models experience mass loss through winds, this only occurs once the stars become red giants \citep{Meynet2006}. Therefore, our results are robust against this effect, particularly with respect to the final masses of black holes, which would form via direct collapse \citep{Heger2003}. We did not include the effect of dynamical friction in our models presented here. Doing this will lead to a faster sinking of massive stars into the cluster core and subsequently may increase the total number of collisions. This may ultimately increase the mass of the MMO.

We have also neglected the stellar feedback during AMUSE simulations which could increase the gas temperature and prevent further star formation. However, previous studies show that the SN feedback is unable to expel the gas from massive embedded star clusters \citep{Krause2016,Grudic2019,Lahen25}. Moreover, star formation can still continue in such system despite pre-SN feedback as the gas forms a disc in the cluster core \citep{Lahen25}. Therefore, we expect that our results would not change significantly by such effects.

Perhaps one of the most surprising results of this study is that the continuous insertion of stars during the simulation (to mimic star formation) delays the core-collapse of the star clusters. This effect is related to the numerical modeling as we assume all new born stars to be in virial equilibrium. This must be considered in future models. One potential solution to mitigate this is to relax the assumption of virial equilibrium when inserting new stars and instead assume some degree of sub-viriality. For our models, this would likely result in clusters that can reach core-collapse within 2 Myr, triggering a runaway collision growth of the MMO and increasing the final masses of the BHs. In this regard, our results likely represent a lower limit on the mass of the BHs.

We evolved our simulations only up to 2 Myrs, as stellar evolution is not included, and thus we are unable to reliably track the fate of the remnants that form at this stage. However, we can assume that each MMO evolves into a black hole of the same mass, with further growth possible through tidal disruption events (TDEs). \citet{Sakurai2019} explored the role of TDEs during BH growth in dense star clusters and found that the TDE rate is

\begin{equation}
    \dot{N}_{\rm TDE} \approx 0.3\ {\rm Myr^{-1}} \left( \frac{M_{\rm IMBH}}{1000\ {\rm M}_\odot} \right)^2.
\end{equation}

In our model with less favorable conditions, we have $M_{\rm BH}\sim 400$~M$_\odot$, implying $\sim2$ TDE per Myr. For our model with $M_{\rm BH}\sim 5000$~M$_\odot$, we get $\sim$ 7 TDE per Myr. In the most optimistic scenario, where  100~M$_\odot$ stars are completely accreted, we might expect an increase of 200 to 700~M$_\odot$ in the mass of these BHs. Thus, our results likely represent a lower limit for the final BH masses.

An important aspect to consider is the retention of BHs in the star clusters, as subsequent BH mergers can result in high-velocity recoil kicks. We estimate the recoil velocities by assuming that the two MMOs in each cluster will merge. These objects tend to form binary systems, and their typical masses are presented in Table \ref{tab:binary_mmo}. Our estimates are based on the fitting formulae provided by \citet{Merritt2004}.

We begin by considering our less massive clusters, which contain a total mass of $4\times10^5$~M$_\odot$ and a corresponding escape velocity of $\sim76$~km~s$^{-1}$. In these clusters, the second MMO lies in the PISN mass range, so we consider a merger between BHs with masses of 420 and 140 M$_\odot$. The recoil velocity ranges from $29$ to $429$~km~s$^{-1}$, indicating that BH ejection is possible in this case. For our clusters with a total mass of $6\times10^5$~M$_\odot$, the escape velocity is 108~km~s$^{-1}$. The recoil velocities range from 10 to 235~km~s$^{-1}$ for the pair with 1000 and 754~M$_\odot$, 24 to 352~km~s$^{-1}$ for the pair with 591 and 329~M$_\odot$, and 3 to 68~km~s$^{-1}$ for a pair with 140 and 2317~M$_\odot$. In this case, BH retention is almost certain for one pair, while remains uncertain for the other two. For our clusters with a total mass of~$1\times10^6$~M$_\odot$, the escape velocity is 132~km~s$^{-1}$. The recoil velocities range from 2 to 27~km~s$^{-1}$ for the pair with 3950 and 140~M$_\odot$, 26 to 398~km~s$^{-1}$ for a pair with 532 and 140~M$_\odot$, and 4 to 69~km~s$^{-1}$ for a pair with 2317 and 140~M$_\odot$. In this case, it is highly likely that these clusters will retain their MBHs. Next, we consider the clusters with rapid SF, which have a stellar mass of $4\times10^5$~M$_\odot$, and an escape velocity of $76$~km~s$^{-1}$. The recoil velocities range from 5 to 88~km~s$^{-1}$ for a pair with 4227 and 335~M$_\odot$, 1 to 20~km~s$^{-1}$ for a pair with 4699 and 140~M$_\odot$, and 1 to 20~km~s$^{-1}$ for a pair with 4685 and 140~M$_\odot$. Again, it is very likely that these clusters will retain their massive BHs.

Ideally, one would need to follow the dynamical evolution of these BHs and their spins over longer timescales, as in \citet{Rantala24}. All of these BHs could grow further through TDEs and if the new stars are initialized in a sub-virial state, as discussed above, the final BH masses could be a factor of $\sim10$ larger. This increase in mass would drastically reduce the recoil kick velocity, ensuring that all of these BHs would most likely be retained in their clusters. In general, our simple estimates suggest that the most massive clusters are highly likely to retain their BHs as they merge with other significantly less massive BHs.

\begin{acknowledgements}
   We gratefully acknowledge the Federal Ministry of Education and Research and the state governments of Chile for supporting this project as part of the funding of the Kultrun cluster hosted at the Departamento de Astronomía, Universidad de Concepción.
   BR acknowledges funding through ANID (CONICYT-PFCHA/Doctorado acuerdo bilateral DAAD/62180013), DAAD (funding program number 57451854), the International Max Planck Research School for Astronomy and Cosmic Physics at the University of Heidelberg (IMPRS-HD), and the support by the European Research Council via ERC Consolidator grant KETJU (no. 818930). DRGS thanks for funding via the  Alexander von Humboldt - Foundation, Bonn, Germany and the ANID BASAL project  FB210003.
\end{acknowledgements}

\bibliographystyle{bibtex/aa} 
\bibliography{bibliography} 

\begin{thebibliography}{126}
\expandafter\ifx\csname natexlab\endcsname\relax\def\natexlab#1{#1}\fi

\bibitem[{{Aarseth}(1999)}]{Nbody_Sverre}
{Aarseth}, S.~J. 1999, \href{http://dx.doi.org/10.1086/316455}{\color{magenta}The Publications of the Astronomical Society of the Pacific}, \href{https://ui.adsabs.harvard.edu/abs/1999PASP..111.1333A}{111, 1333}

\bibitem[{{Abel} {et~al.}(1997){Abel}, {Anninos}, {Zhang}, \& {Norman}}]{Abel97}
{Abel}, T., {Anninos}, P., {Zhang}, Y., \& {Norman}, M.~L. 1997, \href{http://dx.doi.org/10.1016/S1384-1076(97)00010-9}{\color{magenta}New Astronomy}, \href{http://adsabs.harvard.edu/abs/1997NewA....2..181A}{2, 181}

\bibitem[{{Adamo} {et~al.}(2024){Adamo}, {Bradley}, {Vanzella}, {Claeyssens}, {Welch}, {Diego}, {Mahler}, {Oguri}, {Sharon}, {Abdurro'uf}, {Hsiao}, {Xu}, {Messa}, {Lassen}, {Zackrisson}, {Brammer}, {Coe}, {Kokorev}, {Ricotti}, {Zitrin}, {Fujimoto}, {Inoue}, {Resseguier}, {Rigby}, {Jim{\'e}nez-Teja}, {Windhorst}, {Hashimoto}, \& {Tamura}}]{Adamo24}
{Adamo}, A., {Bradley}, L.~D., {Vanzella}, E., {et~al.} 2024, \href{http://dx.doi.org/10.1038/s41586-024-07703-7}{\color{magenta}\nat}, \href{https://ui.adsabs.harvard.edu/abs/2024Natur.632..513A}{632, 513}

\bibitem[{{Alvarez} {et~al.}(2009){Alvarez}, {Wise}, \& {Abel}}]{Alvarez2009}
{Alvarez}, M.~A., {Wise}, J.~H., \& {Abel}, T. 2009, \href{http://dx.doi.org/10.1088/0004-637X/701/2/L133}{\color{magenta}\apjl}, \href{https://ui.adsabs.harvard.edu/abs/2009ApJ...701L.133A}{701, L133}

\bibitem[{{Arca Sedda} {et~al.}(2024){Arca Sedda}, {Kamlah}, {Spurzem}, {Giersz}, {Berczik}, {Rastello}, {Iorio}, {Mapelli}, {Gatto}, \& {Grebel}}]{ArcaSedda24}
{Arca Sedda}, M., {Kamlah}, A. W.~H., {Spurzem}, R., {et~al.} 2024, \href{http://dx.doi.org/10.1093/mnras/stad3952}{\color{magenta}\mnras}, \href{https://ui.adsabs.harvard.edu/abs/2024MNRAS.528.5119A}{528, 5119}

\bibitem[{{Arca Sedda} {et~al.}(2023){Arca Sedda}, {Kamlah}, {Spurzem}, {Rizzuto}, {Naab}, {Giersz}, \& {Berczik}}]{ArcaSedda23}
{Arca Sedda}, M., {Kamlah}, A. W.~H., {Spurzem}, R., {et~al.} 2023, \href{http://dx.doi.org/10.1093/mnras/stad2292}{\color{magenta}\mnras}, \href{https://ui.adsabs.harvard.edu/abs/2023MNRAS.526..429A}{526, 429}

\bibitem[{{Ba{\~n}ados} {et~al.}(2021){Ba{\~n}ados}, {Mazzucchelli}, {Momjian}, {Eilers}, {Wang}, {Schindler}, {Connor}, {Andika}, {Barth}, {Carilli}, {Davies}, {Decarli}, {Fan}, {Farina}, {Hennawi}, {Pensabene}, {Stern}, {Venemans}, {Wenzl}, \& {Yang}}]{Banados2021}
{Ba{\~n}ados}, E., {Mazzucchelli}, C., {Momjian}, E., {et~al.} 2021, \href{http://dx.doi.org/10.3847/1538-4357/abe239}{\color{magenta}\apj}, \href{https://ui.adsabs.harvard.edu/abs/2021ApJ...909...80B}{909, 80}

\bibitem[{{Ba{\~n}ados} {et~al.}(2018){Ba{\~n}ados}, {Venemans}, {Mazzucchelli}, {Farina}, {Walter}, {Wang}, {Decarli}, {Stern}, {Fan}, {Davies}, {Hennawi}, {Simcoe}, {Turner}, {Rix}, {Yang}, {Kelson}, {Rudie}, \& {Winters}}]{Banados2018}
{Ba{\~n}ados}, E., {Venemans}, B.~P., {Mazzucchelli}, C., {et~al.} 2018, \href{http://dx.doi.org/10.1038/nature25180}{\color{magenta}\nat}, \href{https://ui.adsabs.harvard.edu/abs/2018Natur.553..473B}{553, 473}

\bibitem[{{Behroozi} {et~al.}(2013){Behroozi}, {Wechsler}, \& {Wu}}]{RS}
{Behroozi}, P.~S., {Wechsler}, R.~H., \& {Wu}, H.-Y. 2013, \href{http://dx.doi.org/10.1088/0004-637X/762/2/109}{\color{magenta}\apj}, \href{https://ui.adsabs.harvard.edu/abs/2013ApJ...762..109B}{762, 109}

\bibitem[{{Binney} \& {Tremaine}(2008)}]{BinneyTremaine2008}
{Binney}, J. \& {Tremaine}, S. 2008, {Galactic Dynamics: Second Edition}

\bibitem[{{Boekholt} {et~al.}(2018){Boekholt}, {Schleicher}, {Fellhauer}, {Klessen}, {Reinoso}, {Stutz}, \& {Haemmerl{\'e}}}]{Boekholt2018}
{Boekholt}, T.~C.~N., {Schleicher}, D.~R.~G., {Fellhauer}, M., {et~al.} 2018, \href{http://dx.doi.org/10.1093/mnras/sty208}{\color{magenta}\mnras}, \href{https://ui.adsabs.harvard.edu/abs/2018MNRAS.476..366B}{476, 366}

\bibitem[{{Bogd{\'a}n} {et~al.}(2024){Bogd{\'a}n}, {Goulding}, {Natarajan}, {Kov{\'a}cs}, {Tremblay}, {Chadayammuri}, {Volonteri}, {Kraft}, {Forman}, {Jones}, {Churazov}, \& {Zhuravleva}}]{Bogdan24}
{Bogd{\'a}n}, {\'A}., {Goulding}, A.~D., {Natarajan}, P., {et~al.} 2024, \href{http://dx.doi.org/10.1038/s41550-023-02111-9}{\color{magenta}Nature Astronomy}, \href{https://ui.adsabs.harvard.edu/abs/2024NatAs...8..126B}{8, 126}

\bibitem[{{Bovino} {et~al.}(2014){Bovino}, {Latif}, {Grassi}, \& {Schleicher}}]{Bovino2014}
{Bovino}, S., {Latif}, M.~A., {Grassi}, T., \& {Schleicher}, D.~R.~G. 2014, \href{http://dx.doi.org/10.1093/mnras/stu714}{\color{magenta}\mnras}, \href{https://ui.adsabs.harvard.edu/abs/2014MNRAS.441.2181B}{441, 2181}

\bibitem[{{Bryan} {et~al.}(2014){Bryan}, {Norman}, {O'Shea}, {Abel}, {Wise}, {Turk}, {Reynolds}, {Collins}, {Wang}, {Skillman}, {Smith}, {Harkness}, {Bordner}, {Kim}, {Kuhlen}, {Xu}, {Goldbaum}, {Hummels}, {Kritsuk}, {Tasker}, {Skory}, {Simpson}, {Hahn}, {Oishi}, {So}, {Zhao}, {Cen}, {Li}, \& {Enzo Collaboration}}]{ENZO}
{Bryan}, G.~L., {Norman}, M.~L., {O'Shea}, B.~W., {et~al.} 2014, \href{http://dx.doi.org/10.1088/0067-0049/211/2/19}{\color{magenta}\apjs}, \href{https://ui.adsabs.harvard.edu/abs/2014ApJS..211...19B}{211, 19}

\bibitem[{{Chon} {et~al.}(2018){Chon}, {Hosokawa}, \& {Yoshida}}]{Chon2018}
{Chon}, S., {Hosokawa}, T., \& {Yoshida}, N. 2018, \href{http://dx.doi.org/10.1093/mnras/sty086}{\color{magenta}\mnras}, \href{https://ui.adsabs.harvard.edu/abs/2018MNRAS.475.4104C}{475, 4104}

\bibitem[{{Chon} \& {Omukai}(2020)}]{Chon2020}
{Chon}, S. \& {Omukai}, K. 2020, \href{http://dx.doi.org/10.1093/mnras/staa863}{\color{magenta}\mnras}, \href{https://ui.adsabs.harvard.edu/abs/2020MNRAS.494.2851C}{494, 2851}

\bibitem[{{Chon} \& {Omukai}(2024)}]{Chon24}
{Chon}, S. \& {Omukai}, K. 2024, \href{https://ui.adsabs.harvard.edu/abs/2024arXiv241214900C}{\href{http://dx.doi.org/10.48550/arXiv.2412.14900}{\color{magenta}arXiv e-prints}, arXiv:2412.14900}

\bibitem[{{Davies} {et~al.}(2011){Davies}, {Miller}, \& {Bellovary}}]{Davies2011}
{Davies}, M.~B., {Miller}, M.~C., \& {Bellovary}, J.~M. 2011, \href{http://dx.doi.org/10.1088/2041-8205/740/2/L42}{\color{magenta}\apjl}, \href{https://ui.adsabs.harvard.edu/abs/2011ApJ...740L..42D}{740, L42}

\bibitem[{{Escala}(2021)}]{Escala2021}
{Escala}, A. 2021, \href{http://dx.doi.org/10.3847/1538-4357/abd93c}{\color{magenta}\apj}, \href{https://ui.adsabs.harvard.edu/abs/2021ApJ...908...57E}{908, 57}

\bibitem[{{Fan} {et~al.}(2006){Fan}, {Strauss}, {Richards}, {Hennawi}, {Becker}, {White}, {Diamond-Stanic}, {Donley}, {Jiang}, {Kim}, {Vestergaard}, {Young}, {Gunn}, {Lupton}, {Knapp}, {Schneider}, {Brandt}, {Bahcall}, {Barentine}, {Brinkmann}, {Brewington}, {Fukugita}, {Harvanek}, {Kleinman}, {Krzesinski}, {Long}, {Neilsen}, {Nitta}, {Snedden}, \& {Voges}}]{Fan2006}
{Fan}, X., {Strauss}, M.~A., {Richards}, G.~T., {et~al.} 2006, \href{http://dx.doi.org/10.1086/500296}{\color{magenta}\aj}, \href{https://ui.adsabs.harvard.edu/abs/2006AJ....131.1203F}{131, 1203}

\bibitem[{{Federrath} {et~al.}(2011){Federrath}, {Sur}, {Schleicher}, {Banerjee}, \& {Klessen}}]{Fed11}
{Federrath}, C., {Sur}, S., {Schleicher}, D. R.~G., {Banerjee}, R., \& {Klessen}, R.~S. 2011, \href{http://dx.doi.org/10.1088/0004-637X/731/1/62}{\color{magenta}\apj}, \href{https://ui.adsabs.harvard.edu/abs/2011ApJ...731...62F}{731, 62}

\bibitem[{{Fujii} {et~al.}(2007){Fujii}, {Iwasawa}, {Funato}, \& {Makino}}]{Fujii2007}
{Fujii}, M., {Iwasawa}, M., {Funato}, Y., \& {Makino}, J. 2007, \href{http://dx.doi.org/10.1093/pasj/59.6.1095}{\color{magenta}\pasj}, \href{https://ui.adsabs.harvard.edu/abs/2007PASJ...59.1095F}{59, 1095}

\bibitem[{{Furtak} {et~al.}(2024){Furtak}, {Labb{\'e}}, {Zitrin}, {Greene}, {Dayal}, {Chemerynska}, {Kokorev}, {Miller}, {Goulding}, {de Graaff}, {Bezanson}, {Brammer}, {Cutler}, {Leja}, {Pan}, {Price}, {Wang}, {Weaver}, {Whitaker}, {Atek}, {Bogd{\'a}n}, {Charlot}, {Curtis-Lake}, {van Dokkum}, {Endsley}, {Feldmann}, {Fudamoto}, {Fujimoto}, {Glazebrook}, {Juneau}, {Marchesini}, {Maseda}, {Nelson}, {Oesch}, {Plat}, {Setton}, {Stark}, \& {Williams}}]{Furt23}
{Furtak}, L.~J., {Labb{\'e}}, I., {Zitrin}, A., {et~al.} 2024, \href{http://dx.doi.org/10.1038/s41586-024-07184-8}{\color{magenta}\nat}, \href{https://ui.adsabs.harvard.edu/abs/2024Natur.628...57F}{628, 57}

\bibitem[{{Gaete} {et~al.}(2024){Gaete}, {Schleicher}, {Lupi}, {Reinoso}, {Fellhauer}, \& {Vergara}}]{Gaete2024}
{Gaete}, B., {Schleicher}, D.~R.~G., {Lupi}, A., {et~al.} 2024, \href{http://dx.doi.org/10.1051/0004-6361/202450770}{\color{magenta}\aap}, \href{https://ui.adsabs.harvard.edu/abs/2024A&A...690A.378G}{690, A378}

\bibitem[{{Giersz} {et~al.}(2024){Giersz}, {Askar}, {Hypki}, {Hong}, {Wiktorowicz}, \& {Hellstrom}}]{Giersz2024}
{Giersz}, M., {Askar}, A., {Hypki}, A., {et~al.} 2024, \href{https://ui.adsabs.harvard.edu/abs/2024arXiv241106421G}{\href{http://dx.doi.org/10.48550/arXiv.2411.06421}{\color{magenta}arXiv e-prints}, arXiv:2411.06421}

\bibitem[{{Glebbeek} {et~al.}(2013){Glebbeek}, {Gaburov}, {Portegies Zwart}, \& {Pols}}]{Glebbeek2013}
{Glebbeek}, E., {Gaburov}, E., {Portegies Zwart}, S., \& {Pols}, O.~R. 2013, \href{http://dx.doi.org/10.1093/mnras/stt1268}{\color{magenta}\mnras}, \href{https://ui.adsabs.harvard.edu/abs/2013MNRAS.434.3497G}{434, 3497}

\bibitem[{{Goulding} {et~al.}(2023){Goulding}, {Greene}, {Setton}, {Labbe}, {Bezanson}, {Miller}, {Atek}, {Bogd{\'a}n}, {Brammer}, {Chemerynska}, {Cutler}, {Dayal}, {Fudamoto}, {Fujimoto}, {Furtak}, {Kokorev}, {Khullar}, {Leja}, {Marchesini}, {Natarajan}, {Nelson}, {Oesch}, {Pan}, {Papovich}, {Price}, {van Dokkum}, {Wang}, {Weaver}, {Whitaker}, \& {Zitrin}}]{Gould23}
{Goulding}, A.~D., {Greene}, J.~E., {Setton}, D.~J., {et~al.} 2023, \href{http://dx.doi.org/10.3847/2041-8213/acf7c5}{\color{magenta}\apjl}, \href{https://ui.adsabs.harvard.edu/abs/2023ApJ...955L..24G}{955, L24}

\bibitem[{{Greene} {et~al.}(2024){Greene}, {Labbe}, {Goulding}, {Furtak}, {Chemerynska}, {Kokorev}, {Dayal}, {Volonteri}, {Williams}, {Wang}, {Setton}, {Burgasser}, {Bezanson}, {Atek}, {Brammer}, {Cutler}, {Feldmann}, {Fujimoto}, {Glazebrook}, {de Graaff}, {Khullar}, {Leja}, {Marchesini}, {Maseda}, {Matthee}, {Miller}, {Naidu}, {Nanayakkara}, {Oesch}, {Pan}, {Papovich}, {Price}, {van Dokkum}, {Weaver}, {Whitaker}, \& {Zitrin}}]{Greene24}
{Greene}, J.~E., {Labbe}, I., {Goulding}, A.~D., {et~al.} 2024, \href{http://dx.doi.org/10.3847/1538-4357/ad1e5f}{\color{magenta}\apj}, \href{https://ui.adsabs.harvard.edu/abs/2024ApJ...964...39G}{964, 39}

\bibitem[{{Greif} {et~al.}(2008){Greif}, {Johnson}, {Klessen}, \& {Bromm}}]{Greif2008}
{Greif}, T.~H., {Johnson}, J.~L., {Klessen}, R.~S., \& {Bromm}, V. 2008, \href{http://dx.doi.org/10.1111/j.1365-2966.2008.13326.x}{\color{magenta}\mnras}, \href{https://ui.adsabs.harvard.edu/abs/2008MNRAS.387.1021G}{387, 1021}

\bibitem[{Grudić {et~al.}(2018)Grudić, Hopkins, Quataert, \& Murray}]{Grudic2019}
Grudić, M.~Y., Hopkins, P.~F., Quataert, E., \& Murray, N. 2018, \href{http://dx.doi.org/10.1093/mnras/sty3386}{\color{magenta}Monthly Notices of the Royal Astronomical Society}, 483, 5548

\bibitem[{{Habouzit} {et~al.}(2016){Habouzit}, {Volonteri}, {Latif}, {Dubois}, \& {Peirani}}]{Habouzit2016}
{Habouzit}, M., {Volonteri}, M., {Latif}, M., {Dubois}, Y., \& {Peirani}, S. 2016, \href{http://dx.doi.org/10.1093/mnras/stw1924}{\color{magenta}\mnras}, \href{https://ui.adsabs.harvard.edu/abs/2016MNRAS.463..529H}{463, 529}

\bibitem[{{Hahn} \& {Abel}(2011)}]{Hahn11}
{Hahn}, O. \& {Abel}, T. 2011, \href{http://dx.doi.org/10.1111/j.1365-2966.2011.18820.x}{\color{magenta}\mnras}, \href{https://ui.adsabs.harvard.edu/abs/2011MNRAS.415.2101H}{415, 2101}

\bibitem[{{Heger} {et~al.}(2003){Heger}, {Fryer}, {Woosley}, {Langer}, \& {Hartmann}}]{Heger2003}
{Heger}, A., {Fryer}, C.~L., {Woosley}, S.~E., {Langer}, N., \& {Hartmann}, D.~H. 2003, \href{http://dx.doi.org/10.1086/375341}{\color{magenta}\apj}, \href{https://ui.adsabs.harvard.edu/abs/2003ApJ...591..288H}{591, 288}

\bibitem[{{Heggie}(1975)}]{Heggie75}
{Heggie}, D.~C. 1975, \href{http://dx.doi.org/10.1093/mnras/173.3.729}{\color{magenta}\mnras}, \href{https://ui.adsabs.harvard.edu/abs/1975MNRAS.173..729H}{173, 729}

\bibitem[{{Hills}(1975)}]{Hills75}
{Hills}, J.~G. 1975, \href{http://dx.doi.org/10.1086/111815}{\color{magenta}\aj}, \href{https://ui.adsabs.harvard.edu/abs/1975AJ.....80..809H}{80, 809}

\bibitem[{{Hirano} {et~al.}(2017){Hirano}, {Hosokawa}, {Yoshida}, \& {Kuiper}}]{Hirano2017}
{Hirano}, S., {Hosokawa}, T., {Yoshida}, N., \& {Kuiper}, R. 2017, \href{http://dx.doi.org/10.1126/science.aai9119}{\color{magenta}Science}, \href{https://ui.adsabs.harvard.edu/abs/2017Sci...357.1375H}{357, 1375}

\bibitem[{{Hosokawa} \& {Omukai}(2009)}]{Hosokawa09}
{Hosokawa}, T. \& {Omukai}, K. 2009, \href{http://dx.doi.org/10.1088/0004-637X/703/2/1810}{\color{magenta}\apj}, \href{https://ui.adsabs.harvard.edu/abs/2009ApJ...703.1810H}{703, 1810}

\bibitem[{{Hosokawa} {et~al.}(2012){Hosokawa}, {Omukai}, \& {Yorke}}]{Hosokawa12}
{Hosokawa}, T., {Omukai}, K., \& {Yorke}, H.~W. 2012, \href{http://dx.doi.org/10.1088/0004-637X/756/1/93}{\color{magenta}\apj}, \href{https://ui.adsabs.harvard.edu/abs/2012ApJ...756...93H}{756, 93}

\bibitem[{{Hosokawa} {et~al.}(2013){Hosokawa}, {Yorke}, {Inayoshi}, {Omukai}, \& {Yoshida}}]{Hosokawa13}
{Hosokawa}, T., {Yorke}, H.~W., {Inayoshi}, K., {Omukai}, K., \& {Yoshida}, N. 2013, \href{http://dx.doi.org/10.1088/0004-637X/778/2/178}{\color{magenta}\apj}, \href{https://ui.adsabs.harvard.edu/abs/2013ApJ...778..178H}{778, 178}

\bibitem[{{Johnson} \& {Bromm}(2007)}]{JohnsonBromm2007}
{Johnson}, J.~L. \& {Bromm}, V. 2007, \href{http://dx.doi.org/10.1111/j.1365-2966.2006.11275.x}{\color{magenta}\mnras}, \href{https://ui.adsabs.harvard.edu/abs/2007MNRAS.374.1557J}{374, 1557}

\bibitem[{{Kamlah} {et~al.}(2022){Kamlah}, {Leveque}, {Spurzem}, {Arca Sedda}, {Askar}, {Banerjee}, {Berczik}, {Giersz}, {Hurley}, {Belloni}, {K{\"u}hmichel}, \& {Wang}}]{Kamlah22}
{Kamlah}, A.~W.~H., {Leveque}, A., {Spurzem}, R., {et~al.} 2022, \href{http://dx.doi.org/10.1093/mnras/stab3748}{\color{magenta}\mnras}, \href{https://ui.adsabs.harvard.edu/abs/2022MNRAS.511.4060K}{511, 4060}

\bibitem[{{Katz} {et~al.}(2015){Katz}, {Sijacki}, \& {Haehnelt}}]{Katz2015}
{Katz}, H., {Sijacki}, D., \& {Haehnelt}, M.~G. 2015, \href{http://dx.doi.org/10.1093/mnras/stv1048}{\color{magenta}\mnras}, \href{https://ui.adsabs.harvard.edu/abs/2015MNRAS.451.2352K}{451, 2352}

\bibitem[{{King}(2024)}]{King24}
{King}, A. 2024, \href{http://dx.doi.org/10.1093/mnras/stae1171}{\color{magenta}\mnras}, \href{https://ui.adsabs.harvard.edu/abs/2024MNRAS.531..550K}{531, 550}

\bibitem[{{Kocevski} {et~al.}(2024){Kocevski}, {Finkelstein}, {Barro}, {Taylor}, {Calabr{\`o}}, {Laloux}, {Buchner}, {Trump}, {Leung}, {Yang}, {Dickinson}, {P{\'e}rez-Gonz{\'a}lez}, {Pacucci}, {Inayoshi}, {Somerville}, {McGrath}, {Akins}, {Bagley}, {Bisigello}, {Bowler}, {Carnall}, {Casey}, {Cheng}, {Cleri}, {Costantin}, {Cullen}, {Davis}, {Donnan}, {Dunlop}, {Ellis}, {Ferguson}, {Fujimoto}, {Fontana}, {Giavalisco}, {Grazian}, {Grogin}, {Hathi}, {Hirschmann}, {Huertas-Company}, {Holwerda}, {Illingworth}, {Juneau}, {Kartaltepe}, {Koekemoer}, {Li}, {Lucas}, {Magee}, {Mason}, {McLeod}, {McLure}, {Napolitano}, {Papovich}, {Pirzkal}, {Rodighiero}, {Santini}, {Wilkins}, \& {Yung}}]{Koc24}
{Kocevski}, D.~D., {Finkelstein}, S.~L., {Barro}, G., {et~al.} 2024, \href{https://ui.adsabs.harvard.edu/abs/2024arXiv240403576K}{\href{http://dx.doi.org/10.48550/arXiv.2404.03576}{\color{magenta}arXiv e-prints}, arXiv:2404.03576}

\bibitem[{{Kokorev} {et~al.}(2024){Kokorev}, {Caputi}, {Greene}, {Dayal}, {Trebitsch}, {Cutler}, {Fujimoto}, {Labb{\'e}}, {Miller}, {Iani}, {Navarro-Carrera}, \& {Rinaldi}}]{Koko24}
{Kokorev}, V., {Caputi}, K.~I., {Greene}, J.~E., {et~al.} 2024, \href{http://dx.doi.org/10.3847/1538-4357/ad4265}{\color{magenta}\apj}, \href{https://ui.adsabs.harvard.edu/abs/2024ApJ...968...38K}{968, 38}

\bibitem[{{Kokorev} {et~al.}(2023){Kokorev}, {Fujimoto}, {Labbe}, {Greene}, {Bezanson}, {Dayal}, {Nelson}, {Atek}, {Brammer}, {Caputi}, {Chemerynska}, {Cutler}, {Feldmann}, {Fudamoto}, {Furtak}, {Goulding}, {de Graaff}, {Leja}, {Marchesini}, {Miller}, {Nanayakkara}, {Oesch}, {Pan}, {Price}, {Setton}, {Smit}, {Stefanon}, {Wang}, {Weaver}, {Whitaker}, {Williams}, \& {Zitrin}}]{Kok23}
{Kokorev}, V., {Fujimoto}, S., {Labbe}, I., {et~al.} 2023, \href{http://dx.doi.org/10.3847/2041-8213/ad037a}{\color{magenta}\apjl}, \href{https://ui.adsabs.harvard.edu/abs/2023ApJ...957L...7K}{957, L7}

\bibitem[{{Kov{\'a}cs} {et~al.}(2024){Kov{\'a}cs}, {Bogd{\'a}n}, {Natarajan}, {Werner}, {Azadi}, {Volonteri}, {Tremblay}, {Chadayammuri}, {Forman}, {Jones}, \& {Kraft}}]{Kovacs24}
{Kov{\'a}cs}, O.~E., {Bogd{\'a}n}, {\'A}., {Natarajan}, P., {et~al.} 2024, \href{http://dx.doi.org/10.3847/2041-8213/ad391f}{\color{magenta}\apjl}, \href{https://ui.adsabs.harvard.edu/abs/2024ApJ...965L..21K}{965, L21}

\bibitem[{{Krause} {et~al.}(2016){Krause}, {Charbonnel}, {Bastian}, \& {Diehl}}]{Krause2016}
{Krause}, M. G.~H., {Charbonnel}, C., {Bastian}, N., \& {Diehl}, R. 2016, \href{http://dx.doi.org/10.1051/0004-6361/201526685}{\color{magenta}\aap}, \href{https://ui.adsabs.harvard.edu/abs/2016A&A...587A..53K}{587, A53}

\bibitem[{{Kroupa} {et~al.}(2020){Kroupa}, {Subr}, {Jerabkova}, \& {Wang}}]{Kroupa2020}
{Kroupa}, P., {Subr}, L., {Jerabkova}, T., \& {Wang}, L. 2020, \href{http://dx.doi.org/10.1093/mnras/staa2276}{\color{magenta}\mnras}, \href{https://ui.adsabs.harvard.edu/abs/2020MNRAS.498.5652K}{498, 5652}

\bibitem[{{Krti{\v{c}}ka} \& {Kub{\'a}t}(2006)}]{Krtivka2006}
{Krti{\v{c}}ka}, J. \& {Kub{\'a}t}, J. 2006, \href{http://dx.doi.org/10.1051/0004-6361:20053289}{\color{magenta}\aap}, \href{https://ui.adsabs.harvard.edu/abs/2006A&A...446.1039K}{446, 1039}

\bibitem[{{Lah{\'e}n} {et~al.}(2025){Lah{\'e}n}, {Naab}, {Rantala}, \& {Partmann}}]{Lahen25}
{Lah{\'e}n}, N., {Naab}, T., {Rantala}, A., \& {Partmann}, C. 2025, \href{https://ui.adsabs.harvard.edu/abs/2025arXiv250418620L}{\href{http://dx.doi.org/10.48550/arXiv.2504.18620}{\color{magenta}arXiv e-prints}, arXiv:2504.18620}

\bibitem[{{Lambrides} {et~al.}(2024{\natexlab{a}}){Lambrides}, {Chiaberge}, {Long}, {Liu}, {Akins}, {Ptak}, {Andika}, {Capetti}, {Casey}, {Champagne}, {Chworowsky}, {Clarke}, {Cooper}, {Ding}, {Dong}, {Faisst}, {Forman}, {Franco}, {Gillman}, {Gozaliasl}, {Hall}, {Harish}, {Hayward}, {Hirschmann}, {Hutchison}, {Jahnke}, {Jin}, {Kartaltepe}, {Kleiner}, {Koekemoer}, {Kokorev}, {Manning}, {Martin}, {McKinney}, {Norman}, {Nyland}, {Onoue}, {Robertson}, {Shuntov}, {Silverman}, {Stiavelli}, {Trakhtenbrot}, {Vardoulaki}, {Zavala}, {Allen}, {Ilbert}, {McCracken}, {Paquereau}, {Rhodes}, \& {Toft}}]{Lamb23}
{Lambrides}, E., {Chiaberge}, M., {Long}, A.~S., {et~al.} 2024{\natexlab{a}}, \href{http://dx.doi.org/10.3847/2041-8213/ad11ee}{\color{magenta}\apjl}, \href{https://ui.adsabs.harvard.edu/abs/2024ApJ...961L..25L}{961, L25}

\bibitem[{{Lambrides} {et~al.}(2024{\natexlab{b}}){Lambrides}, {Garofali}, {Larson}, {Ptak}, {Chiaberge}, {Long}, {Hutchison}, {Norman}, {McKinney}, {Akins}, {Berg}, {Chisholm}, {Civano}, {Cloonan}, {Endsley}, {Faisst}, {Gilli}, {Gillman}, {Hirschmann}, {Kartaltepe}, {Kocevski}, {Kokorev}, {Pacucci}, {Richardson}, {Stiavelli}, \& {Whalen}}]{Lambrides24}
{Lambrides}, E., {Garofali}, K., {Larson}, R., {et~al.} 2024{\natexlab{b}}, \href{https://ui.adsabs.harvard.edu/abs/2024arXiv240913047L}{\href{http://dx.doi.org/10.48550/arXiv.2409.13047}{\color{magenta}arXiv e-prints}, arXiv:2409.13047}

\bibitem[{{Larson} {et~al.}(2023){Larson}, {Finkelstein}, {Kocevski}, {Hutchison}, {Trump}, {Arrabal Haro}, {Bromm}, {Cleri}, {Dickinson}, {Fujimoto}, {Kartaltepe}, {Koekemoer}, {Papovich}, {Pirzkal}, {Tacchella}, {Zavala}, {Bagley}, {Behroozi}, {Champagne}, {Cole}, {Jung}, {Morales}, {Yang}, {Zhang}, {Zitrin}, {Amor{\'\i}n}, {Burgarella}, {Casey}, {Ch{\'a}vez Ortiz}, {Cox}, {Chworowsky}, {Fontana}, {Gawiser}, {Grazian}, {Grogin}, {Harish}, {Hathi}, {Hirschmann}, {Holwerda}, {Juneau}, {Leung}, {Lucas}, {McGrath}, {P{\'e}rez-Gonz{\'a}lez}, {Rigby}, {Seill{\'e}}, {Simons}, {de La Vega}, {Weiner}, {Wilkins}, {Yung}, \& {Ceers Team}}]{Lars23}
{Larson}, R.~L., {Finkelstein}, S.~L., {Kocevski}, D.~D., {et~al.} 2023, \href{http://dx.doi.org/10.3847/2041-8213/ace619}{\color{magenta}\apjl}, \href{https://ui.adsabs.harvard.edu/abs/2023ApJ...953L..29L}{953, L29}

\bibitem[{{Latif} \& {Khochfar}(2020)}]{latif20b}
{Latif}, M.~A. \& {Khochfar}, S. 2020, \href{http://dx.doi.org/10.1093/mnras/staa2218}{\color{magenta}\mnras}, \href{https://ui.adsabs.harvard.edu/abs/2020MNRAS.497.3761L}{497, 3761}

\bibitem[{{Latif} {et~al.}(2021){Latif}, {Khochfar}, {Schleicher}, \& {Whalen}}]{Latif2021}
{Latif}, M.~A., {Khochfar}, S., {Schleicher}, D., \& {Whalen}, D.~J. 2021, \href{http://dx.doi.org/10.1093/mnras/stab2708}{\color{magenta}\mnras}, \href{https://ui.adsabs.harvard.edu/abs/2021MNRAS.508.1756L}{508, 1756}

\bibitem[{{Latif} {et~al.}(2014){Latif}, {Schleicher}, {Bovino}, {Grassi}, \& {Spaans}}]{LAtif2014}
{Latif}, M.~A., {Schleicher}, D.~R.~G., {Bovino}, S., {Grassi}, T., \& {Spaans}, M. 2014, \href{http://dx.doi.org/10.1088/0004-637X/792/1/78}{\color{magenta}\apj}, \href{https://ui.adsabs.harvard.edu/abs/2014ApJ...792...78L}{792, 78}

\bibitem[{{Latif} {et~al.}(2016){Latif}, {Schleicher}, \& {Hartwig}}]{Latif2016}
{Latif}, M.~A., {Schleicher}, D.~R.~G., \& {Hartwig}, T. 2016, \href{http://dx.doi.org/10.1093/mnras/stw297}{\color{magenta}\mnras}, \href{https://ui.adsabs.harvard.edu/abs/2016MNRAS.458..233L}{458, 233}

\bibitem[{{Latif} {et~al.}(2013){Latif}, {Schleicher}, {Schmidt}, \& {Niemeyer}}]{Latif13c}
{Latif}, M.~A., {Schleicher}, D.~R.~G., {Schmidt}, W., \& {Niemeyer}, J. 2013, \href{http://dx.doi.org/10.1093/mnras/stt834}{\color{magenta}\mnras}, \href{http://adsabs.harvard.edu/abs/2013MNRAS.433.1607L}{433, 1607}

\bibitem[{{Latif} \& {Volonteri}(2015)}]{Latif2015}
{Latif}, M.~A. \& {Volonteri}, M. 2015, \href{http://dx.doi.org/10.1093/mnras/stv1337}{\color{magenta}\mnras}, \href{https://ui.adsabs.harvard.edu/abs/2015MNRAS.452.1026L}{452, 1026}

\bibitem[{{Latif} {et~al.}(2022{\natexlab{a}}){Latif}, {Whalen}, \& {Khochfar}}]{L22}
{Latif}, M.~A., {Whalen}, D., \& {Khochfar}, S. 2022{\natexlab{a}}, \href{http://dx.doi.org/10.3847/1538-4357/ac3916}{\color{magenta}\apj}, \href{https://ui.adsabs.harvard.edu/abs/2022ApJ...925...28L}{925, 28}

\bibitem[{{Latif} {et~al.}(2022{\natexlab{b}}){Latif}, {Whalen}, {Khochfar}, {Herrington}, \& {Woods}}]{L22N}
{Latif}, M.~A., {Whalen}, D.~J., {Khochfar}, S., {Herrington}, N.~P., \& {Woods}, T.~E. 2022{\natexlab{b}}, \href{http://dx.doi.org/10.1038/s41586-022-04813-y}{\color{magenta}\nat}, \href{https://ui.adsabs.harvard.edu/abs/2022Natur.607...48L}{607, 48}

\bibitem[{{Liempi} {et~al.}(2025){Liempi}, {Schleicher}, {Benson}, {Escala}, \& {Vergara}}]{Liempi2025}
{Liempi}, M., {Schleicher}, D.~R.~G., {Benson}, A., {Escala}, A., \& {Vergara}, M.~C. 2025, \href{http://dx.doi.org/10.1051/0004-6361/202451672}{\color{magenta}\aap}, \href{https://ui.adsabs.harvard.edu/abs/2025A&A...694A..42L}{694, A42}

\bibitem[{{Lupi} {et~al.}(2014){Lupi}, {Colpi}, {Devecchi}, {Galanti}, \& {Volonteri}}]{Lupi2014}
{Lupi}, A., {Colpi}, M., {Devecchi}, B., {Galanti}, G., \& {Volonteri}, M. 2014, \href{http://dx.doi.org/10.1093/mnras/stu1120}{\color{magenta}\mnras}, \href{https://ui.adsabs.harvard.edu/abs/2014MNRAS.442.3616L}{442, 3616}

\bibitem[{{Lupi} {et~al.}(2024){Lupi}, {Trinca}, {Volonteri}, {Dotti}, \& {Mazzucchelli}}]{Lupi2024}
{Lupi}, A., {Trinca}, A., {Volonteri}, M., {Dotti}, M., \& {Mazzucchelli}, C. 2024, \href{http://dx.doi.org/10.1051/0004-6361/202451249}{\color{magenta}\aap}, \href{https://ui.adsabs.harvard.edu/abs/2024A&A...689A.128L}{689, A128}

\bibitem[{{Maio} {et~al.}(2007){Maio}, {Dolag}, {Ciardi}, \& {Tornatore}}]{Maio2007}
{Maio}, U., {Dolag}, K., {Ciardi}, B., \& {Tornatore}, L. 2007, \href{http://dx.doi.org/10.1111/j.1365-2966.2007.12016.x}{\color{magenta}\mnras}, \href{https://ui.adsabs.harvard.edu/abs/2007MNRAS.379..963M}{379, 963}

\bibitem[{{Maiolino} {et~al.}(2025){Maiolino}, {Risaliti}, {Signorini}, {Trefoloni}, {Juod{\v{z}}balis}, {Scholtz}, {{\"U}bler}, {D'Eugenio}, {Carniani}, {Fabian}, {Ji}, {Mazzolari}, {Bertola}, {Brusa}, {Bunker}, {Charlot}, {Comastri}, {Cresci}, {DeCoursey}, {Egami}, {Fiore}, {Gilli}, {Perna}, {Tacchella}, \& {Venturi}}]{Mai24}
{Maiolino}, R., {Risaliti}, G., {Signorini}, M., {et~al.} 2025, \href{http://dx.doi.org/10.1093/mnras/staf359}{\color{magenta}\mnras} \href{https://ui.adsabs.harvard.edu/abs/2025MNRAS.tmp..337M}{[\eprint[arXiv]{2405.00504}]}

\bibitem[{{Maiolino} {et~al.}(2024){Maiolino}, {Scholtz}, {Witstok}, {Carniani}, {D'Eugenio}, {de Graaff}, {{\"U}bler}, {Tacchella}, {Curtis-Lake}, {Arribas}, {Bunker}, {Charlot}, {Chevallard}, {Curti}, {Looser}, {Maseda}, {Rawle}, {Rodr{\'\i}guez del Pino}, {Willott}, {Egami}, {Eisenstein}, {Hainline}, {Robertson}, {Williams}, {Willmer}, {Baker}, {Boyett}, {DeCoursey}, {Fabian}, {Helton}, {Ji}, {Jones}, {Kumari}, {Laporte}, {Nelson}, {Perna}, {Sandles}, {Shivaei}, \& {Sun}}]{Maio23a}
{Maiolino}, R., {Scholtz}, J., {Witstok}, J., {et~al.} 2024, \href{http://dx.doi.org/10.1038/s41586-024-07052-5}{\color{magenta}\nat}, \href{https://ui.adsabs.harvard.edu/abs/2024Natur.627...59M}{627, 59}

\bibitem[{{Matthee} {et~al.}(2024){Matthee}, {Naidu}, {Brammer}, {Chisholm}, {Eilers}, {Goulding}, {Greene}, {Kashino}, {Labbe}, {Lilly}, {Mackenzie}, {Oesch}, {Weibel}, {Wuyts}, {Xiao}, {Bordoloi}, {Bouwens}, {van Dokkum}, {Illingworth}, {Kramarenko}, {Maseda}, {Mason}, {Meyer}, {Nelson}, {Reddy}, {Shivaei}, {Simcoe}, \& {Yue}}]{Mat24}
{Matthee}, J., {Naidu}, R.~P., {Brammer}, G., {et~al.} 2024, \href{http://dx.doi.org/10.3847/1538-4357/ad2345}{\color{magenta}\apj}, \href{https://ui.adsabs.harvard.edu/abs/2024ApJ...963..129M}{963, 129}

\bibitem[{{Mayer} {et~al.}(2025){Mayer}, {van Donkelaar}, {Messa}, {Capelo}, \& {Adamo}}]{Mayer25}
{Mayer}, L., {van Donkelaar}, F., {Messa}, M., {Capelo}, P.~R., \& {Adamo}, A. 2025, \href{http://dx.doi.org/10.3847/2041-8213/adadfe}{\color{magenta}\apjl}, \href{https://ui.adsabs.harvard.edu/abs/2025ApJ...981L..28M}{981, L28}

\bibitem[{{McGreer} \& {Bryan}(2008)}]{McGreer2008}
{McGreer}, I.~D. \& {Bryan}, G.~L. 2008, \href{http://dx.doi.org/10.1086/590530}{\color{magenta}\apj}, \href{https://ui.adsabs.harvard.edu/abs/2008ApJ...685....8M}{685, 8}

\bibitem[{{McMillan} \& {Hut}(1996)}]{McMillan96}
{McMillan}, S. L.~W. \& {Hut}, P. 1996, \href{http://dx.doi.org/10.1086/177610}{\color{magenta}\apj}, \href{https://ui.adsabs.harvard.edu/abs/1996ApJ...467..348M}{467, 348}

\bibitem[{Merritt {et~al.}(2004)Merritt, Milosavljević, Favata, Hughes, \& Holz}]{Merritt2004}
Merritt, D., Milosavljević, M., Favata, M., Hughes, S.~A., \& Holz, D.~E. 2004, \href{http://dx.doi.org/10.1086/421551}{\color{magenta}\apjl}, 607, L9

\bibitem[{{Meynet} {et~al.}(2006){Meynet}, {Ekstr{\"o}m}, \& {Maeder}}]{Meynet2006}
{Meynet}, G., {Ekstr{\"o}m}, S., \& {Maeder}, A. 2006, \href{http://dx.doi.org/10.1051/0004-6361:20053070}{\color{magenta}\aap}, \href{https://ui.adsabs.harvard.edu/abs/2006A&A...447..623M}{447, 623}

\bibitem[{{Mirocha} {et~al.}(2012){Mirocha}, {Skory}, {Burns}, \& {Wise}}]{Mirocha12}
{Mirocha}, J., {Skory}, S., {Burns}, J.~O., \& {Wise}, J.~H. 2012, \href{http://dx.doi.org/10.1088/0004-637X/756/1/94}{\color{magenta}\apj}, \href{https://ui.adsabs.harvard.edu/abs/2012ApJ...756...94M}{756, 94}

\bibitem[{{Mortlock} {et~al.}(2011){Mortlock}, {Warren}, {Venemans}, {Patel}, {Hewett}, {McMahon}, {Simpson}, {Theuns}, {Gonz{\'a}les-Solares}, {Adamson}, {Dye}, {Hambly}, {Hirst}, {Irwin}, {Kuiper}, {Lawrence}, \& {R{\"o}ttgering}}]{Mortlock2011}
{Mortlock}, D.~J., {Warren}, S.~J., {Venemans}, B.~P., {et~al.} 2011, \href{http://dx.doi.org/10.1038/nature10159}{\color{magenta}\nat}, \href{https://ui.adsabs.harvard.edu/abs/2011Natur.474..616M}{474, 616}

\bibitem[{{Mukherjee} {et~al.}(2025){Mukherjee}, {Zhou}, {Chen}, {Di Carlo}, \& {Di Matteo}}]{Mukherjee24}
{Mukherjee}, D., {Zhou}, Y., {Chen}, N., {Di Carlo}, U.~N., \& {Di Matteo}, T. 2025, \href{http://dx.doi.org/10.3847/1538-4357/adb1b0}{\color{magenta}\apj}, \href{https://ui.adsabs.harvard.edu/abs/2025ApJ...981..203M}{981, 203}

\bibitem[{{Naidu} {et~al.}(2025){Naidu}, {Matthee}, {Katz}, {de Graaff}, {Oesch}, {Smith}, {Greene}, {Brammer}, {Weibel}, {Hviding}, {Chisholm}, {Labb\textbackslash'e}, {Simcoe}, {Witten}, {Atek}, {Baggen}, {Belli}, {Bezanson}, {Boogaard}, {Bose}, {Covelo-Paz}, {Dayal}, {Fudamoto}, {Furtak}, {Giovinazzo}, {Goulding}, {Gronke}, {Heintz}, {Hirschmann}, {Illingworth}, {Inoue}, {Johnson}, {Leja}, {Leonova}, {McConachie}, {Maseda}, {Natarajan}, {Nelson}, {Setton}, {Shivaei}, {Sobral}, {Stefanon}, {Tacchella}, {Toft}, {Torralba}, {van Dokkum}, {van der Wel}, {Volonteri}, {Walter}, {Wang}, \& {Watson}}]{Naidu25}
{Naidu}, R.~P., {Matthee}, J., {Katz}, H., {et~al.} 2025, \href{https://ui.adsabs.harvard.edu/abs/2025arXiv250316596N}{\href{http://dx.doi.org/10.48550/arXiv.2503.16596}{\color{magenta}arXiv e-prints}, arXiv:2503.16596}

\bibitem[{{Napolitano} {et~al.}(2024){Napolitano}, {Castellano}, {Pentericci}, {Vignali}, {Gilli}, {Fontana}, {Santini}, {Treu}, {Calabr{\`o}}, {Llerena}, {Piconcelli}, {Zappacosta}, {Mascia}, {Bergamini}, {Bakx}, {Dickinson}, {Glazebrook}, {Henry}, {Leethochawalit}, {Mazzolari}, {Merlin}, {Morishita}, {Nanayakkara}, {Paris}, {Puccetti}, {Roberts-Borsani}, {Rojas Ruiz}, {Vanzella}, {Vito}, {Vulcani}, {Wang}, {Yoon}, \& {Zavala}}]{Napolitano24}
{Napolitano}, L., {Castellano}, M., {Pentericci}, L., {et~al.} 2024, \href{https://ui.adsabs.harvard.edu/abs/2024arXiv241018763N}{\href{http://dx.doi.org/10.48550/arXiv.2410.18763}{\color{magenta}arXiv e-prints}, arXiv:2410.18763}

\bibitem[{{Nitadori} \& {Aarseth}(2012)}]{Nitadori2012}
{Nitadori}, K. \& {Aarseth}, S.~J. 2012, \href{http://dx.doi.org/10.1111/j.1365-2966.2012.21227.x}{\color{magenta}\mnras}, \href{https://ui.adsabs.harvard.edu/abs/2012MNRAS.424..545N}{424, 545}

\bibitem[{{O'Brennan} {et~al.}(2025){O'Brennan}, {Regan}, {Brennan}, {McCaffrey}, {Wise}, {Visbal}, \& {Norman}}]{OBrennan25}
{O'Brennan}, H., {Regan}, J.~A., {Brennan}, J., {et~al.} 2025, \href{https://ui.adsabs.harvard.edu/abs/2025arXiv250200574O}{\href{http://dx.doi.org/10.48550/arXiv.2502.00574}{\color{magenta}arXiv e-prints}, arXiv:2502.00574}

\bibitem[{{Omukai} {et~al.}(2008){Omukai}, {Schneider}, \& {Haiman}}]{Omukai2008}
{Omukai}, K., {Schneider}, R., \& {Haiman}, Z. 2008, \href{http://dx.doi.org/10.1086/591636}{\color{magenta}\apj}, \href{https://ui.adsabs.harvard.edu/abs/2008ApJ...686..801O}{686, 801}

\bibitem[{{Onoue} {et~al.}(2019){Onoue}, {Kashikawa}, {Matsuoka}, {Kato}, {Izumi}, {Nagao}, {Strauss}, {Harikane}, {Imanishi}, {Ito}, {Iwasawa}, {Kawaguchi}, {Lee}, {Noboriguchi}, {Suh}, {Tanaka}, \& {Toba}}]{Onoue2019}
{Onoue}, M., {Kashikawa}, N., {Matsuoka}, Y., {et~al.} 2019, \href{http://dx.doi.org/10.3847/1538-4357/ab29e9}{\color{magenta}\apj}, \href{https://ui.adsabs.harvard.edu/abs/2019ApJ...880...77O}{880, 77}

\bibitem[{{Pacucci} {et~al.}(2017){Pacucci}, {Natarajan}, {Volonteri}, {Cappelluti}, \& {Urry}}]{Pacucci2017}
{Pacucci}, F., {Natarajan}, P., {Volonteri}, M., {Cappelluti}, N., \& {Urry}, C.~M. 2017, \href{http://dx.doi.org/10.3847/2041-8213/aa9aea}{\color{magenta}\apjl}, \href{https://ui.adsabs.harvard.edu/abs/2017ApJ...850L..42P}{850, L42}

\bibitem[{{Partmann} {et~al.}(2025){Partmann}, {Naab}, {Lah{\'e}n}, {Rantala}, {Hirschmann}, {Hislop}, {Petersson}, \& {Johansson}}]{Partmann25}
{Partmann}, C., {Naab}, T., {Lah{\'e}n}, N., {et~al.} 2025, \href{http://dx.doi.org/10.1093/mnras/staf002}{\color{magenta}\mnras}, \href{https://ui.adsabs.harvard.edu/abs/2025MNRAS.537..956P}{537, 956}

\bibitem[{{Patrick} {et~al.}(2023){Patrick}, {Whalen}, {Latif}, \& {Elford}}]{pat23}
{Patrick}, S.~J., {Whalen}, D.~J., {Latif}, M.~A., \& {Elford}, J.~S. 2023, \href{http://dx.doi.org/10.1093/mnras/stad1179}{\color{magenta}\mnras}, \href{https://ui.adsabs.harvard.edu/abs/2023MNRAS.522.3795P}{522, 3795}

\bibitem[{{Pelupessy} {et~al.}(2013){Pelupessy}, {van Elteren}, {de Vries}, {McMillan}, {Drost}, \& {Portegies Zwart}}]{AMUSE_Pelupessy13}
{Pelupessy}, F.~I., {van Elteren}, A., {de Vries}, N., {et~al.} 2013, \href{http://dx.doi.org/10.1051/0004-6361/201321252}{\color{magenta}\aap}, \href{https://ui.adsabs.harvard.edu/abs/2013A&A...557A..84P}{557, A84}

\bibitem[{{Planck Collaboration} {et~al.}(2016){Planck Collaboration}, {Ade}, {Aghanim}, {Arnaud}, {Ashdown}, {Aumont}, {Baccigalupi}, {Banday}, {Barreiro}, {Bartlett}, {Bartolo}, {Battaner}, {Battye}, {Benabed}, {Beno{\^\i}t}, {Benoit-L{\'e}vy}, {Bernard}, {Bersanelli}, {Bielewicz}, {Bock}, {Bonaldi}, {Bonavera}, {Bond}, {Borrill}, {Bouchet}, {Boulanger}, {Bucher}, {Burigana}, {Butler}, {Calabrese}, {Cardoso}, {Catalano}, {Challinor}, {Chamballu}, {Chary}, {Chiang}, {Chluba}, {Christensen}, {Church}, {Clements}, {Colombi}, {Colombo}, {Combet}, {Coulais}, {Crill}, {Curto}, {Cuttaia}, {Danese}, {Davies}, {Davis}, {de Bernardis}, {de Rosa}, {de Zotti}, {Delabrouille}, {D{\'e}sert}, {Di Valentino}, {Dickinson}, {Diego}, {Dolag}, {Dole}, {Donzelli}, {Dor{\'e}}, {Douspis}, {Ducout}, {Dunkley}, {Dupac}, {Efstathiou}, {Elsner}, {En{\ss}lin}, {Eriksen}, {Farhang}, {Fergusson}, {Finelli}, {Forni}, {Frailis}, {Fraisse}, {Franceschi}, {Frejsel}, {Galeotta}, {Galli}, {Ganga}, {Gauthier}, {Gerbino}, {Ghosh}, {Giard},
  {Giraud-H{\'e}raud}, {Giusarma}, {Gjerl{\o}w}, {Gonz{\'a}lez-Nuevo}, {G{\'o}rski}, {Gratton}, {Gregorio}, {Gruppuso}, {Gudmundsson}, {Hamann}, {Hansen}, {Hanson}, {Harrison}, {Helou}, {Henrot-Versill{\'e}}, {Hern{\'a}ndez-Monteagudo}, {Herranz}, {Hildebrandt}, {Hivon}, {Hobson}, {Holmes}, {Hornstrup}, {Hovest}, {Huang}, {Huffenberger}, {Hurier}, {Jaffe}, {Jaffe}, {Jones}, {Juvela}, {Keih{\"a}nen}, {Keskitalo}, {Kisner}, {Kneissl}, {Knoche}, {Knox}, {Kunz}, {Kurki-Suonio}, {Lagache}, {L{\"a}hteenm{\"a}ki}, {Lamarre}, {Lasenby}, {Lattanzi}, {Lawrence}, {Leahy}, {Leonardi}, {Lesgourgues}, {Levrier}, {Lewis}, {Liguori}, {Lilje}, {Linden-V{\o}rnle}, {L{\'o}pez-Caniego}, {Lubin}, {Mac{\'\i}as-P{\'e}rez}, {Maggio}, {Maino}, {Mandolesi}, {Mangilli}, {Marchini}, {Maris}, {Martin}, {Martinelli}, {Mart{\'\i}nez-Gonz{\'a}lez}, {Masi}, {Matarrese}, {McGehee}, {Meinhold}, {Melchiorri}, {Melin}, {Mendes}, {Mennella}, {Migliaccio}, {Millea}, {Mitra}, {Miville-Desch{\^e}nes}, {Moneti}, {Montier}, {Morgante}, {Mortlock},
  {Moss}, {Munshi}, {Murphy}, {Naselsky}, {Nati}, {Natoli}, {Netterfield}, {N{\o}rgaard-Nielsen}, {Noviello}, {Novikov}, {Novikov}, {Oxborrow}, {Paci}, {Pagano}, {Pajot}, {Paladini}, {Paoletti}, {Partridge}, {Pasian}, {Patanchon}, {Pearson}, {Perdereau}, {Perotto}, {Perrotta}, {Pettorino}, {Piacentini}, {Piat}, {Pierpaoli}, {Pietrobon}, {Plaszczynski}, {Pointecouteau}, {Polenta}, {Popa}, {Pratt}, {Pr{\'e}zeau}, {Prunet}, {Puget}, {Rachen}, {Reach}, {Rebolo}, {Reinecke}, {Remazeilles}, {Renault}, {Renzi}, {Ristorcelli}, {Rocha}, {Rosset}, {Rossetti}, {Roudier}, {Rouill{\'e} d'Orfeuil}, {Rowan-Robinson}, {Rubi{\~n}o-Mart{\'\i}n}, {Rusholme}, {Said}, {Salvatelli}, {Salvati}, {Sandri}, {Santos}, {Savelainen}, {Savini}, {Scott}, {Seiffert}, {Serra}, {Shellard}, {Spencer}, {Spinelli}, {Stolyarov}, {Stompor}, {Sudiwala}, {Sunyaev}, {Sutton}, {Suur-Uski}, {Sygnet}, {Tauber}, {Terenzi}, {Toffolatti}, {Tomasi}, {Tristram}, {Trombetti}, {Tucci}, {Tuovinen}, {T{\"u}rler}, {Umana}, {Valenziano}, {Valiviita}, {Van Tent},
  {Vielva}, {Villa}, {Wade}, {Wandelt}, {Wehus}, {White}, {White}, {Wilkinson}, {Yvon}, {Zacchei}, \& {Zonca}}]{Planck16}
{Planck Collaboration}, {Ade}, P.~A.~R., {Aghanim}, N., {et~al.} 2016, \href{http://dx.doi.org/10.1051/0004-6361/201525830}{\color{magenta}\aap}, \href{https://ui.adsabs.harvard.edu/abs/2016A&A...594A..13P}{594, A13}

\bibitem[{{Plummer}(1911)}]{Plummer1911}
{Plummer}, H.~C. 1911, \href{http://dx.doi.org/10.1093/mnras/71.5.460}{\color{magenta}\mnras}, \href{https://ui.adsabs.harvard.edu/abs/1911MNRAS..71..460P}{71, 460}

\bibitem[{{Portegies Zwart} \& {McMillan}(2018)}]{Portegies2018}
{Portegies Zwart}, S. \& {McMillan}, S. 2018, {Astrophysical Recipes; The art of AMUSE}

\bibitem[{{Portegies Zwart} {et~al.}(2009){Portegies Zwart}, {McMillan}, {Harfst}, {Groen}, {Fujii}, {Nuall{\'a}in}, {Glebbeek}, {Heggie}, {Lombardi}, {Hut}, {Angelou}, {Banerjee}, {Belkus}, {Fragos}, {Fregeau}, {Gaburov}, {Izzard}, {Juri{\'c}}, {Justham}, {Sottoriva}, {Teuben}, {van Bever}, {Yaron}, \& {Zemp}}]{AMUSE_Portegies09}
{Portegies Zwart}, S., {McMillan}, S., {Harfst}, S., {et~al.} 2009, \href{http://dx.doi.org/10.1016/j.newast.2008.10.006}{\color{magenta}\na}, \href{https://ui.adsabs.harvard.edu/abs/2009NewA...14..369P}{14, 369}

\bibitem[{{Portegies Zwart} {et~al.}(2013){Portegies Zwart}, {McMillan}, {van Elteren}, {Pelupessy}, \& {de Vries}}]{AMUSE_Portegies13}
{Portegies Zwart}, S., {McMillan}, S.~L.~W., {van Elteren}, E., {Pelupessy}, I., \& {de Vries}, N. 2013, \href{http://dx.doi.org/10.1016/j.cpc.2012.09.024}{\color{magenta}Computer Physics Communications}, \href{https://ui.adsabs.harvard.edu/abs/2013CoPhC.184..456P}{184, 456}

\bibitem[{{Portegies Zwart} {et~al.}(2004){Portegies Zwart}, {Baumgardt}, {Hut}, {Makino}, \& {McMillan}}]{PortegiesZwart2004}
{Portegies Zwart}, S.~F., {Baumgardt}, H., {Hut}, P., {Makino}, J., \& {McMillan}, S. L.~W. 2004, \href{http://dx.doi.org/10.1038/nature02448}{\color{magenta}\nat}, \href{https://ui.adsabs.harvard.edu/abs/2004Natur.428..724P}{428, 724}

\bibitem[{{Rantala} {et~al.}(2024){Rantala}, {Naab}, \& {Lah{\'e}n}}]{Rantala24}
{Rantala}, A., {Naab}, T., \& {Lah{\'e}n}, N. 2024, \href{http://dx.doi.org/10.1093/mnras/stae1413}{\color{magenta}\mnras}, \href{https://ui.adsabs.harvard.edu/abs/2024MNRAS.531.3770R}{531, 3770}

\bibitem[{{Reed} {et~al.}(2019){Reed}, {Banerji}, {Becker}, {Hewett}, {Martini}, {McMahon}, {Pons}, {Rauch}, {Abbott}, {Allam}, {Annis}, {Avila}, {Bertin}, {Brooks}, {Buckley-Geer}, {Carnero Rosell}, {Carrasco Kind}, {Carretero}, {Castander}, {Cunha}, {D'Andrea}, {da Costa}, {De Vicente}, {Desai}, {Diehl}, {Doel}, {Evrard}, {Flaugher}, {Frieman}, {Garc{\'\i}a-Bellido}, {Gaztanaga}, {Gruen}, {Gschwend}, {Gutierrez}, {Hollowood}, {Honscheid}, {Hoyle}, {James}, {Kuehn}, {Lahav}, {Lima}, {Maia}, {Marshall}, {Miquel}, {Ogand o}, {Plazas}, {Roodman}, {Sanchez}, {Scarpine}, {Schubnell}, {Serrano}, {Sevilla-Noarbe}, {Smith}, {Smith}, {Sobreira}, {Suchyta}, {Swanson}, {Tarle}, {Thomas}, {Tucker}, \& {Vikram}}]{Reed2019}
{Reed}, S.~L., {Banerji}, M., {Becker}, G.~D., {et~al.} 2019, \href{http://dx.doi.org/10.1093/mnras/stz1341}{\color{magenta}\mnras}, \href{https://ui.adsabs.harvard.edu/abs/2019MNRAS.487.1874R}{487, 1874}

\bibitem[{{Regan} \& {Volonteri}(2024)}]{Regan2024}
{Regan}, J. \& {Volonteri}, M. 2024, \href{http://dx.doi.org/10.33232/001c.123239}{\color{magenta}The Open Journal of Astrophysics}, \href{https://ui.adsabs.harvard.edu/abs/2024OJAp....7E..72R}{7, 72}

\bibitem[{{Regan} \& {Downes}(2018)}]{Regan18b}
{Regan}, J.~A. \& {Downes}, T.~P. 2018, \href{http://dx.doi.org/10.1093/mnras/sty134}{\color{magenta}\mnras}, \href{https://ui.adsabs.harvard.edu/abs/2018MNRAS.475.4636R}{475, 4636}

\bibitem[{{Regan} \& {Haehnelt}(2009{\natexlab{a}})}]{Regan2009}
{Regan}, J.~A. \& {Haehnelt}, M.~G. 2009{\natexlab{a}}, \href{http://dx.doi.org/10.1111/j.1365-2966.2009.14579.x}{\color{magenta}\mnras}, \href{https://ui.adsabs.harvard.edu/abs/2009MNRAS.396..343R}{396, 343}

\bibitem[{{Regan} \& {Haehnelt}(2009{\natexlab{b}})}]{Regan2009b}
{Regan}, J.~A. \& {Haehnelt}, M.~G. 2009{\natexlab{b}}, \href{http://dx.doi.org/10.1111/j.1365-2966.2008.14088.x}{\color{magenta}\mnras}, \href{https://ui.adsabs.harvard.edu/abs/2009MNRAS.393..858R}{393, 858}

\bibitem[{{Regan} {et~al.}(2020){Regan}, {Wise}, {Woods}, {Downes}, {O'Shea}, \& {Norman}}]{Regan2020b}
{Regan}, J.~A., {Wise}, J.~H., {Woods}, T.~E., {et~al.} 2020, \href{http://dx.doi.org/10.21105/astro.2008.08090}{\color{magenta}The Open Journal of Astrophysics}, \href{https://ui.adsabs.harvard.edu/abs/2020OJAp....3E..15R}{3, 15}

\bibitem[{{Reinoso} {et~al.}(2023){Reinoso}, {Klessen}, {Schleicher}, {Glover}, \& {Solar}}]{Reinoso23}
{Reinoso}, B., {Klessen}, R.~S., {Schleicher}, D., {Glover}, S. C.~O., \& {Solar}, P. 2023, \href{http://dx.doi.org/10.1093/mnras/stad790}{\color{magenta}\mnras}, \href{https://ui.adsabs.harvard.edu/abs/2023MNRAS.521.3553R}{521, 3553}

\bibitem[{{Reinoso} {et~al.}(2018){Reinoso}, {Schleicher}, {Fellhauer}, {Klessen}, \& {Boekholt}}]{Reinoso2018}
{Reinoso}, B., {Schleicher}, D.~R.~G., {Fellhauer}, M., {Klessen}, R.~S., \& {Boekholt}, T.~C.~N. 2018, \href{http://dx.doi.org/10.1051/0004-6361/201732224}{\color{magenta}\aap}, \href{https://ui.adsabs.harvard.edu/abs/2018A&A...614A..14R}{614, A14}

\bibitem[{{Reinoso} {et~al.}(2020){Reinoso}, {Schleicher}, {Fellhauer}, {Leigh}, \& {Klessen}}]{Reinoso20}
{Reinoso}, B., {Schleicher}, D.~R.~G., {Fellhauer}, M., {Leigh}, N.~W.~C., \& {Klessen}, R.~S. 2020, \href{http://dx.doi.org/10.1051/0004-6361/202037843}{\color{magenta}\aap}, \href{https://ui.adsabs.harvard.edu/abs/2020A&A...639A..92R}{639, A92}

\bibitem[{{Rusakov} {et~al.}(2025){Rusakov}, {Watson}, {Nikopoulos}, {Brammer}, {Gottumukkala}, {Harvey}, {Heintz}, {Nielsen}, {Sim}, {Sneppen}, {Vijayan}, {Adams}, {Austin}, {Conselice}, {Goolsby}, {Toft}, \& {Witstok}}]{Rusakov25}
{Rusakov}, V., {Watson}, D., {Nikopoulos}, G.~P., {et~al.} 2025, \href{https://ui.adsabs.harvard.edu/abs/2025arXiv250316595R}{\href{http://dx.doi.org/10.48550/arXiv.2503.16595}{\color{magenta}arXiv e-prints}, arXiv:2503.16595}

\bibitem[{{Saavedra-Bastidas} {et~al.}(2024){Saavedra-Bastidas}, {Schleicher}, {Klessen}, {Chon}, {Omukai}, {Peters}, {Prole}, {Reinoso}, {Riaz}, \& {Solar}}]{Saavedra2024}
{Saavedra-Bastidas}, J., {Schleicher}, D. R.~G., {Klessen}, R.~S., {et~al.} 2024, \href{http://dx.doi.org/10.1051/0004-6361/202450409}{\color{magenta}\aap}, \href{https://ui.adsabs.harvard.edu/abs/2024A&A...690A.186S}{690, A186}

\bibitem[{{Sakurai} {et~al.}(2019){Sakurai}, {Yoshida}, \& {Fujii}}]{Sakurai2019}
{Sakurai}, Y., {Yoshida}, N., \& {Fujii}, M.~S. 2019, \href{http://dx.doi.org/10.1093/mnras/stz315}{\color{magenta}\mnras}, \href{https://ui.adsabs.harvard.edu/abs/2019MNRAS.484.4665S}{484, 4665}

\bibitem[{{Sakurai} {et~al.}(2017){Sakurai}, {Yoshida}, {Fujii}, \& {Hirano}}]{Sakurai2017}
{Sakurai}, Y., {Yoshida}, N., {Fujii}, M.~S., \& {Hirano}, S. 2017, \href{http://dx.doi.org/10.1093/mnras/stx2044}{\color{magenta}\mnras}, \href{https://ui.adsabs.harvard.edu/abs/2017MNRAS.472.1677S}{472, 1677}

\bibitem[{{Sassano} {et~al.}(2021){Sassano}, {Schneider}, {Valiante}, {Inayoshi}, {Chon}, {Omukai}, {Mayer}, \& {Capelo}}]{Sassano2021}
{Sassano}, F., {Schneider}, R., {Valiante}, R., {et~al.} 2021, \href{http://dx.doi.org/10.1093/mnras/stab1737}{\color{magenta}\mnras}, \href{https://ui.adsabs.harvard.edu/abs/2021MNRAS.506..613S}{506, 613}

\bibitem[{{Schaerer}(2002)}]{Schaerer02}
{Schaerer}, D. 2002, \href{http://dx.doi.org/10.1051/0004-6361:20011619}{\color{magenta}\aap}, \href{http://esoads.eso.org/abs/2002A%26A...382...28S}{382, 28}

\bibitem[{{Schleicher} {et~al.}(2023){Schleicher}, {Reinoso}, \& {Klessen}}]{Schleicher2023}
{Schleicher}, D. R.~G., {Reinoso}, B., \& {Klessen}, R.~S. 2023, \href{http://dx.doi.org/10.1093/mnras/stad807}{\color{magenta}\mnras}, \href{https://ui.adsabs.harvard.edu/abs/2023MNRAS.521.3972S}{521, 3972}

\bibitem[{{Schleicher} {et~al.}(2022){Schleicher}, {Reinoso}, {Latif}, {Klessen}, {Vergara}, {Das}, {Alister}, {D{\'\i}az}, \& {Solar}}]{Schleicher2022}
{Schleicher}, D.~R.~G., {Reinoso}, B., {Latif}, M., {et~al.} 2022, \href{http://dx.doi.org/10.1093/mnras/stac926}{\color{magenta}\mnras}, \href{https://ui.adsabs.harvard.edu/abs/2022MNRAS.512.6192S}{512, 6192}

\bibitem[{{Schneider} {et~al.}(2023){Schneider}, {Valiante}, {Trinca}, {Graziani}, {Volonteri}, \& {Maiolino}}]{Schneeider2023}
{Schneider}, R., {Valiante}, R., {Trinca}, A., {et~al.} 2023, \href{http://dx.doi.org/10.1093/mnras/stad2503}{\color{magenta}\mnras}, \href{https://ui.adsabs.harvard.edu/abs/2023MNRAS.526.3250S}{526, 3250}

\bibitem[{{Smith} {et~al.}(2018){Smith}, {Regan}, {Downes}, {Norman}, {O'Shea}, \& {Wise}}]{Smith2018}
{Smith}, B.~D., {Regan}, J.~A., {Downes}, T.~P., {et~al.} 2018, \href{http://dx.doi.org/10.1093/mnras/sty2103}{\color{magenta}\mnras}, \href{https://ui.adsabs.harvard.edu/abs/2018MNRAS.480.3762S}{480, 3762}

\bibitem[{{Spitzer}(1987)}]{Spitzer1987}
{Spitzer}, L. 1987, {Dynamical evolution of globular clusters}

\bibitem[{{Spurzem}(1999)}]{Superzem_Nbody6pp}
{Spurzem}, R. 1999, \href{http://dx.doi.org/10.48550/arXiv.astro-ph/9906154}{\color{magenta}Journal of Computational and Applied Mathematics}, \href{https://ui.adsabs.harvard.edu/abs/1999JCoAM.109..407S}{109, 407}

\bibitem[{{Tagawa} {et~al.}(2020){Tagawa}, {Haiman}, \& {Kocsis}}]{Tagawa2020}
{Tagawa}, H., {Haiman}, Z., \& {Kocsis}, B. 2020, \href{http://dx.doi.org/10.3847/1538-4357/ab7922}{\color{magenta}\apj}, \href{https://ui.adsabs.harvard.edu/abs/2020ApJ...892...36T}{892, 36}

\bibitem[{{Trinca} {et~al.}(2024){Trinca}, {Valiante}, {Schneider}, {Juod{\v{z}}balis}, {Maiolino}, {Graziani}, {Lupi}, {Natarajan}, {Volonteri}, \& {Zana}}]{Trinca24}
{Trinca}, A., {Valiante}, R., {Schneider}, R., {et~al.} 2024, \href{https://ui.adsabs.harvard.edu/abs/2024arXiv241214248T}{\href{http://dx.doi.org/10.48550/arXiv.2412.14248}{\color{magenta}arXiv e-prints}, arXiv:2412.14248}

\bibitem[{{Vergara} {et~al.}(2023){Vergara}, {Escala}, {Schleicher}, \& {Reinoso}}]{Vergara23}
{Vergara}, M.~C., {Escala}, A., {Schleicher}, D.~R.~G., \& {Reinoso}, B. 2023, \href{http://dx.doi.org/10.1093/mnras/stad1253}{\color{magenta}\mnras}, \href{https://ui.adsabs.harvard.edu/abs/2023MNRAS.522.4224V}{522, 4224}

\bibitem[{{Vergara} {et~al.}(2024){Vergara}, {Schleicher}, {Escala}, {Reinoso}, {Flammini Dotti}, {Kamlah}, {Liempi}, {Hoyer}, {Neumayer}, \& {Spurzem}}]{Vergara2024}
{Vergara}, M.~C., {Schleicher}, D.~R.~G., {Escala}, A., {et~al.} 2024, \href{http://dx.doi.org/10.1051/0004-6361/202449967}{\color{magenta}\aap}, \href{https://ui.adsabs.harvard.edu/abs/2024A&A...689A..34V}{689, A34}

\bibitem[{{Vergara} {et~al.}(2021){Vergara}, {Schleicher}, {Boekholt}, {Reinoso}, {Fellhauer}, {Klessen}, \& {Leigh}}]{Vergara2021}
{Vergara}, M.~Z.~C., {Schleicher}, D.~R.~G., {Boekholt}, T.~C.~N., {et~al.} 2021, \href{http://dx.doi.org/10.1051/0004-6361/202140298}{\color{magenta}\aap}, \href{https://ui.adsabs.harvard.edu/abs/2021A&A...649A.160V}{649, A160}

\bibitem[{{Wang} {et~al.}(2021){Wang}, {Yang}, {Fan}, {Hennawi}, {Barth}, {Banados}, {Bian}, {Boutsia}, {Connor}, {Davies}, {Decarli}, {Eilers}, {Farina}, {Green}, {Jiang}, {Li}, {Mazzucchelli}, {Nanni}, {Schindler}, {Venemans}, {Walter}, {Wu}, \& {Yue}}]{Wang21}
{Wang}, F., {Yang}, J., {Fan}, X., {et~al.} 2021, \href{http://dx.doi.org/10.3847/2041-8213/abd8c6}{\color{magenta}\apjl}, \href{https://ui.adsabs.harvard.edu/abs/2021ApJ...907L...1W}{907, L1}

\bibitem[{{Wang} {et~al.}(2015){Wang}, {Spurzem}, {Aarseth}, {Nitadori}, {Berczik}, {Kouwenhoven}, \& {Naab}}]{Wang_Nbody6ppgpu}
{Wang}, L., {Spurzem}, R., {Aarseth}, S., {et~al.} 2015, \href{http://dx.doi.org/10.1093/mnras/stv817}{\color{magenta}\mnras}, \href{https://ui.adsabs.harvard.edu/abs/2015MNRAS.450.4070W}{450, 4070}

\bibitem[{{Wise} {et~al.}(2019){Wise}, {Regan}, {O'Shea}, {Norman}, {Downes}, \& {Xu}}]{Wise2019}
{Wise}, J.~H., {Regan}, J.~A., {O'Shea}, B.~W., {et~al.} 2019, \href{http://dx.doi.org/10.1038/s41586-019-0873-4}{\color{magenta}\nat}, \href{https://ui.adsabs.harvard.edu/abs/2019Natur.566...85W}{566, 85}

\bibitem[{{Woods} {et~al.}(2017){Woods}, {Heger}, {Whalen}, {Haemmerl{\'e}}, \& {Klessen}}]{tyr17}
{Woods}, T.~E., {Heger}, A., {Whalen}, D.~J., {Haemmerl{\'e}}, L., \& {Klessen}, R.~S. 2017, \href{http://dx.doi.org/10.3847/2041-8213/aa7412}{\color{magenta}\apjl}, \href{http://adsabs.harvard.edu/abs/2017ApJ...842L...6W}{842, L6}

\bibitem[{{Wu} {et~al.}(2015){Wu}, {Wang}, {Fan}, {Yi}, {Zuo}, {Bian}, {Jiang}, {McGreer}, {Wang}, {Yang}, {Yang}, {Thompson}, \& {Beletsky}}]{Wu15Nature}
{Wu}, X.-B., {Wang}, F., {Fan}, X., {et~al.} 2015, \href{http://dx.doi.org/10.1038/nature14241}{\color{magenta}\nat}, \href{https://ui.adsabs.harvard.edu/abs/2015Natur.518..512W}{518, 512}

\bibitem[{{Yoshida} {et~al.}(2007){Yoshida}, {Omukai}, \& {Hernquist}}]{Yoshida2007}
{Yoshida}, N., {Omukai}, K., \& {Hernquist}, L. 2007, \href{http://dx.doi.org/10.1086/522202}{\color{magenta}\apjl}, \href{https://ui.adsabs.harvard.edu/abs/2007ApJ...667L.117Y}{667, L117}

\end{thebibliography}
%

\begin{appendix} 
\section{Orbital parameters that led to stellar collisions}
\label{sec:orb_params_colls}
In this section we present a compilation of orbital parameters for several pairs of stars that collide in our simulations. For each of the collisions in our simulations we searched for the colliding stars in the snapshot just prior to the event.
We then calculated the specific orbital energy of the pair and computed the orbital parameters. Finally, we calculated the pericenter distance $r_p$ of the orbit. For all the systems in which $r_p \leq {\rm R}_1 + {\rm R}_2$, we presume that the orbit derived from this analysis is the orbit that finally led to the stellar collision. We present in Table~\ref{tab:orb_param_coll_set} the orbital parameters for all these systems, as they serve for producing realistic initial conditions for the study of stellar collisions between massive stars in dense star clusters.

   \begin{table*}[h]
      \caption[]{Subset of collisions from our simulations. The first two columns from left to right indicate the mass and radius of the most massive star, followed by the mass and radius of the less massive star. Columns five and six show the distance and magnitude of the relative velocity between the stars. Column 7 is the impact parameter. Column 8 is the pericenter distance in units of the sum of the radii of the stars. Column 9 shows the time until the collision of the stars, as obtained from our simulations. Column 10 presents the kinetic energy of the collision divided by the binding energy of the most massive star. Column 11 indicates the type of orbit that led to the collision.}
         \label{tab:orb_param_coll_set}
         \centering          
         \begin{tabular}{r r r r r r r r r r l}     
         \hline\hline       
m$_1$ \  & R$_1$ \ & m$_2$ \ & R$_2$ \ & d \ \ \ & v \ \ \ \  & b \ \ \ \ & r$_{\rm p}$ \ \ \ & $\Delta t_{ coll}$ & $E_k/E_b$ & Type \\
 $\left[ {\rm M}_\odot \right]$ & [R$_\odot$] & [M$_\odot$] & [R$_\odot$] & [AU] & [km s$^{-1}$] & [AU] & [R$_1$+R$_2$] & [yr] \ & & \\
 \hline
 2860.31 & 90.56 & 737.98 & 41.84 & 585.60 & 78.62 & 16.73 & 0.44 & 20.27 & 0.15 & Binary \\
 543.19 & 35.13 & 216.10 & 20.77 & 135.37 & 130.50 & 4.12 & 0.83 & 3.62 & 0.23 & Hyperbolic \\
 276.21 & 23.89 & 205.61 & 20.19 & 501.44 & 46.73 & 7.36 & 0.67 & 35.49 & 0.47 & Hyperbolic \\
 188.96 & 19.24 & 179.47 & 18.69 & 563.93 & 84.14 & 2.52 & 0.39 & 27.42 & 0.62 & Hyperbolic \\
 173.98 & 18.36 & 11.89 & 3.98 & 720.51 & 32.58 & 0.44 & 0.01 & 80.72 & 0.04 & Hyperbolic \\
169.79 & 18.11 & 78.02 & 11.62 & 1346.44 & 52.02 & 1.20 & 0.06 & 108.87 & 0.27 & Hyperbolic \\
148.23 & 16.76 & 29.64 & 6.70 & 2371.13 & 126.80 & 0.83 & 0.32 & 86.93 & 0.11 & Hyperbolic \\
79.46 & 11.75 & 10.94 & 3.79 & 1167.97 & 19.98 & 4.06 & 0.57 & 220.12 & 0.08 & Hyperbolic \\
37.50 & 7.65 & 7.68 & 3.10 & 1110.80 & 27.30 & 0.32 & 0.02 & 173.65 & 0.11 & Hyperbolic \\
18.60 & 5.13 & 1.88 & 1.39 & 800.18 & 16.16 & 1.66 & 0.66 & 201.31 & 0.05 & Hyperbolic \\
         \hline                  
      \end{tabular}
   \end{table*}

\section{Gas inflow and halo growth}
\label{sec:appB}
\begin{figure}[h]
   \centering
  \includegraphics[width=9cm]{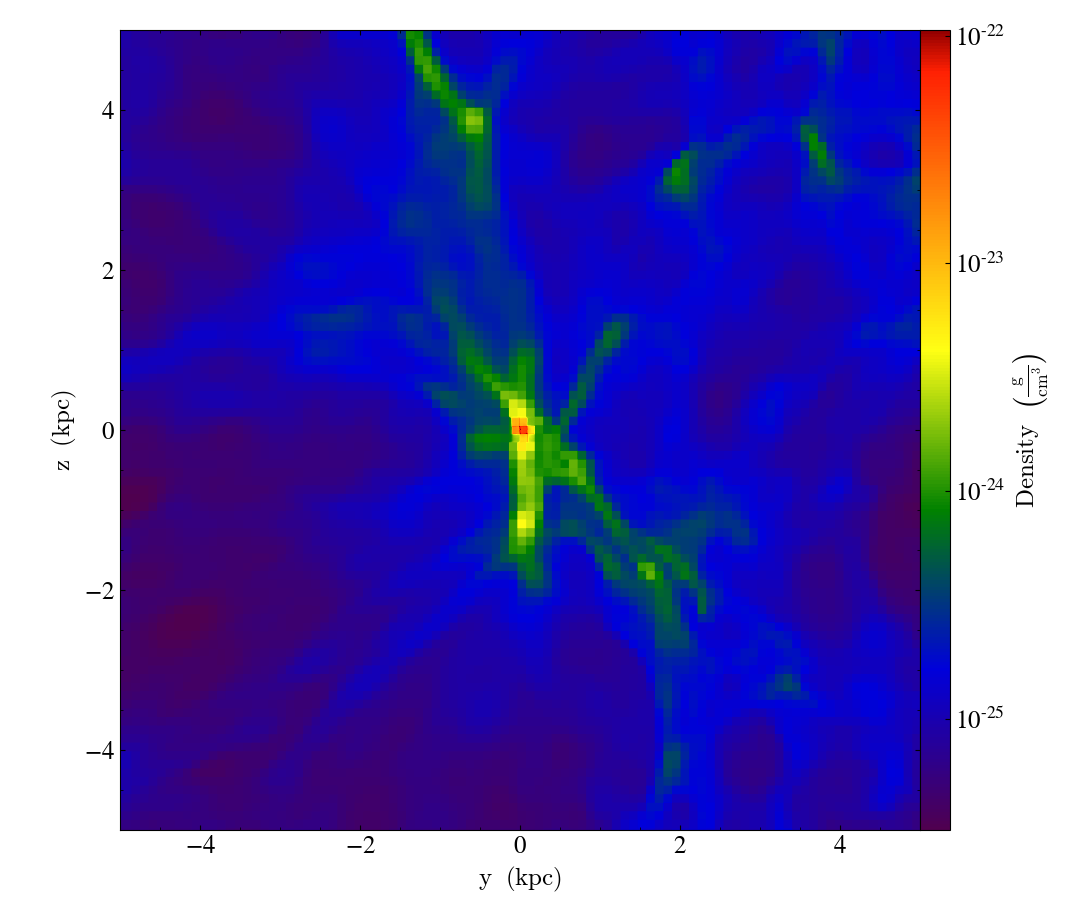}
  \caption{Large scale view of the halo showing its environment and cold dense streams of gas that drive turbulent motions at the center.}
  \label{fig:large_scale_halo}
\end{figure}

\begin{figure}[h]
   \centering
  \includegraphics[width=9cm]{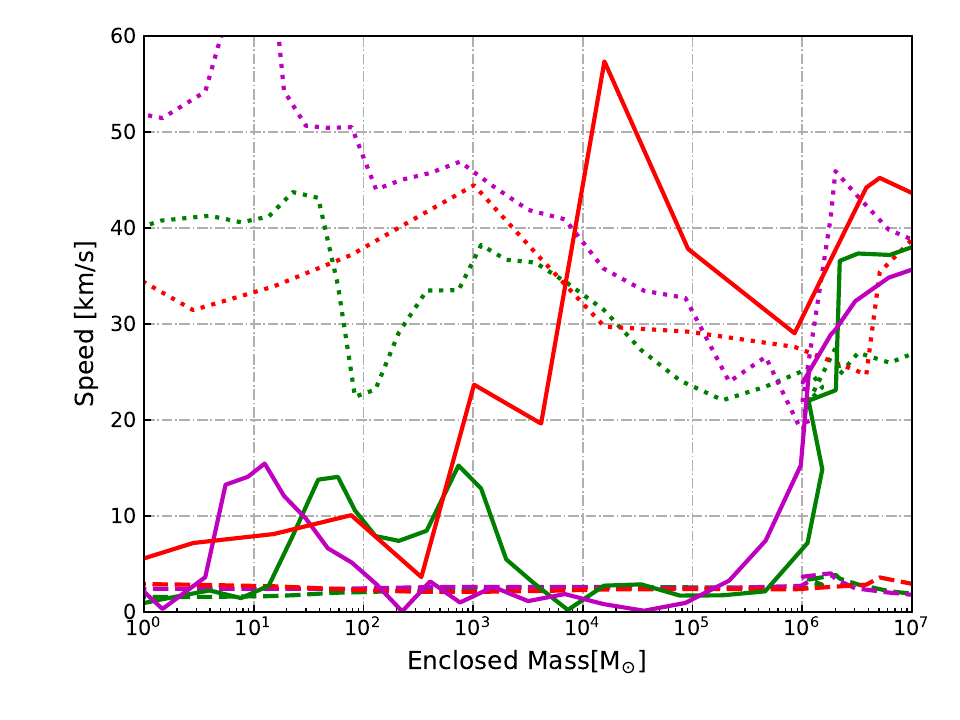}
  \caption{Gas velocity profiles are shown against the enclosed mass. Dotted lines show the turbulent velocities, solid lines the infall velocities and dashed lines  thermal sound speeds. Magenta, green and red colors show time evolution (from earlier to later times), respectively.}
  \label{fig:velocity_profile}
\end{figure}

\begin{figure}[h]
   \centering
  \includegraphics[width=9cm]{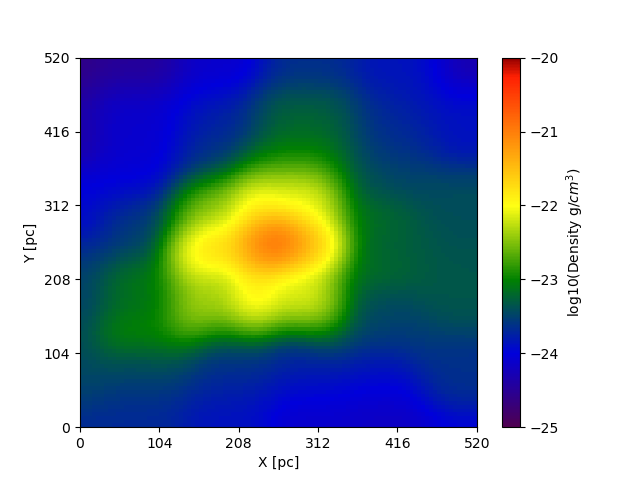}
  \caption{View of the halo on a scale corresponding to the virial radius, which is equal to 260 pc.}
  \label{fig:rvir_scale_halo}
\end{figure}

The gaseous collapse in the minihalo was prevented in a similar fashion to \citet{L22N} and Cold dense gas inflows resulted in rapid gas accretion as shown in Fig.~\ref{fig:large_scale_halo}. Turbulent motions prevented gas infall until the halo grew up to a few times $10^7$~M$_\odot$, and large inflow velocities overcame the turbulent motions leading to catastrophic collapse, see Fig.~\ref{fig:velocity_profile}. We also present a snapshot of the halo on a scale of the virial radius in Fig.~\ref{fig:rvir_scale_halo}

\end{appendix}

\end{document}